\documentclass[12pt,oneside]{article}

\usepackage{epsfig,lscape}
\usepackage{amssymb,amsfonts,amsmath}
\usepackage{rotating}

\usepackage{amsthm}

\def\var{\mathrm{var}}
\def\cov{\mathrm{cov}}
\def\N{\mathrm{N}}

\def\T{ {\mathrm{\scriptscriptstyle T}} }

\newtheorem{pro}{Proposition}
\newtheorem{cor}{Corollary}

\theoremstyle{definition}

\theoremstyle{definition}
\newtheorem{rem}{Remark}

\begin{document}

\begin{titlepage}

\begin{center}
{\Large Analysis of odds, probability, and hazard ratios: From 2 by 2 tables to two-sample survival data}

\vspace{.1in} Zhiqiang Tan\footnotemark[1]

\vspace{.1in}
\today
\end{center}

\footnotetext[1]{Department of Statistics, Rutgers University. Address: 110 Frelinghuysen Road,
Piscataway, NJ 08854. E-mail: ztan@stat.rutgers.edu. The research was supported in part by PCORI grant ME-1511-32740.}

\paragraph{Abstract.} Analysis of 2 by 2 tables and two-sample survival data has been widely used.
Exact calculation is computational intractable for conditional likelihood inference in odds ratio  models with large marginals in 2 by 2 tables,
or partial likelihood inference in Cox's proportional hazards models with considerable tied event times.
Approximate methods are often employed, but their statistical properties have not been formally studied while taking into account the approximation involved.
We develop new methods and theory by constructing suitable estimating functions while leveraging knowledge from conditional or partial likelihood inference.
We propose a weighted Mantel--Haenszel estimator in an odds ratio model such as Cox's discrete-time proportional hazards model.
Moreover, we consider a probability ratio model, and derive as a consistent estimator the Breslow--Peto estimator, which has been regarded as an approximation to partial likelihood estimation in the odds ratio model.
We study both model-based and model-robust variance estimation. For the Breslow--Peto estimator, our new model-based variance estimator is no greater than the
commonly reported variance estimator. We present numerical studies which support the theoretical findings.


\end{titlepage}

\section{Introduction} \label{sect:intro}

Analysis of $2 \times 2$ tables and two-sample survival data has been widely used. The subjects are covered in numerous articles and books
(e.g., Anderson et al.~1993; Breslow \& Day 1980; Cox \& Oaks 1984; Kalbfleisch \& Prentice 1980; McCullagh \& Nelder 1989).
The dominant approach is to use odds ratio models and conditional likelihood inference for handling $2\times 2$ tables,
and Cox's (1972) proportional hazards models and partial likelihood inference for analyzing censored survival data.
The two methods are closely related or, to some extent, equivalent to each other.
In fact, conditional logistic regression can be implemented by
calling a computer routine for fitting Cox's regression model, as seen in the popular R package \texttt{survival} (Therneau 2015).

There are, however, open problems in the existing theory and methods. For Cox's proportional hazards model with survival data, the partial likelihood is
powerful and analytically simple in the absence of tied event times. Large sample theory has been developed using counting processes,
where the event time is commonly assumed to be absolutely continuous, thereby excluding the possibility of tied event times.
But as remarked by Cox (1972), ``Unfortunately it is quite likely in applications that the data will be recorded in a form involving ties."
A possible approach for handling tied data is to use Cox's (1972) discrete-time version of proportional hazards models, which amounts to modeling
odds ratios of hazard probabilities. The associated partial likelihood is conceptually straightforward, but
exact calculation is numerically difficult with a moderate or large number of ties.
Alternatively, various ad hoc approximations to the exact partial likelihood have been proposed (Breslow 1974; Efron 1977; Peto 1972).
It is often said that such approximations could yield satisfactory results with a small number of ties, but this reasoning defeats the very purpose of using
approximations to deal with a relatively large number of ties.
There seems to be no formal theory to justify these approximate methods or study their operating characteristics.
In fact, if the event time is truly discrete, then the estimators of Breslow (1974) and Efron (1977)
would in general be inconsistent  under Cox's discrete-time propositional hazards model.
For discrete-time survival analysis, it has also been proposed to use unconditional maximum likelihood estimation,
either with pooled logistic regression, corresponding to Cox's discrete-time model, or complementary log-log regression induced by
grouping observations under Cox's continuous-time model (Prentice \& Gloeckler 1978). See Allison (1982) for a review.
But such methods seem problematic in the presence of a large number of event times or intervals,
which lead to the same number of nuisance parameters.

There are similar issues in the existing theory and methods for analyzing $2 \times 2$ tables under odds ratio models (Zelen 1971; Breslow 1976).
Although various methods were proposed in the early literature including Mantel--Haenszel estimation (Cochran 1954; Mantel \& Haenszel 1959),
conditional maximum likelihood estimation (Breslow 1981; McCullagh \& Nelder 1989) has been regarded as the ``gold standard" for various reasons
including optimal asymptotic properties (Lindsay 1980) and superior empirical performance (Hauck 1984), as remarked by Breslow \& Cologne (1986).
In particular, conditional likelihood inference is well-behaved with a fixed number of large tables or a large number of sparse tables.
On the other hand, exact calculation for conditional likelihood estimation is numerically intractable for tables with large maginals,
similarly to partial likelihood estimation with a large number of ties.
Approximate methods have been proposed, with supportive numerical evidence (Breslow \& Cologne 1986; McCullagh \& Nelder 1989).
But there seems to be no formal analysis of statistical properties of these methods, while taking into account the approximation involved.

To address the foregoing issues, we develop new methods and theory for analyzing $2\times 2$ tables and two-sample survival data, by constructing suitable estimating functions,
while leveraging knowledge from conditional or partial likelihood inference.
First, in an odds ratio model for $2 \times 2$ tables, we propose a weighted Mantel--Haenszel estimator
by carefully deriving a vector of analytically simple estimating functions from two related angles.
One is to achieve a close approximation to optimal estimating functions in minimizing asymptotic variances when positive responses are rare,
which are often satisfied in applications.
The other is to mimic conditional likelihood estimation in the extreme case where the total number of successes is 1 in each $2\times 2$ table.
The weighted Mantel--Haenszel estimator can be shown to be consistent and asymptotically normal in two asymptotic settings with large tables or many sparse tables,
provided that the odds ratio model is valid.
Moreover, to complement model-based inference, we derive a model-robust variance estimator, which are consistent in both asymptotic settings, while allowing
for possible misspecification of the odds ratio model.
See Buja et al.~(2019) for related discussion of model-robust variance estimation in linear regression.

Seond and perhaps more importantly, we study inference in a probability ratio model, where the log ratio of two success probabilities, instead of their odds ratio, is determined by a linear predictor in each
$2 \times 2$ table. We carefully construct a vector of simple estimating functions, which not only provides a reasonable approximation to optimal estimating functions
in minimizing asymptotic variances, but also coincides with the Brelow--Peto approximation of the conditional likelihood in the odds ratio model.
The resulting estimator, called the Breslow--Peto estimator, can be shown to be consistent and asymptotically normal in two asymptotic settings with large tables or many sparse tables,
provided that the probability ratio model is valid. Moreover, we derive a model-robust variance estimator and a model-based variance estimator. The new model-based variance estimator is shown to
be no greater than the commonly reported model-based variance estimator for the Breslow--Peto estimator, although the two variance estimators are identical
in the special case of a total of one success in each table.

Finally, for two-sample survival analysis, we directly adopt the weighted Mantel--Haenszel estimator in an odds ratio model,
or the Breslow--Peto estimator in a probability ratio model, with $2 \times 2$ tables constructed as usual from risk sets over time.
Both models reduce to Cox's proportional hazards model in the continuous-time limit.
The model-based variance estimators from $2\times 2$ tables can be seen to remain valid due to a martingale argument.
We then derive model-robust variance estimators for the weighted Mantel--Haenszel estimator and the Breslow--Peto estimator.
The latter variance estimator can be shown to coincide with Lin \& Wei's (1989) variance estimator when extended to the Breslow--Peto estimator in the presence of tied event times,
and hence to remain appropriate in Cox's proportional hazards model in continuous time.
As a result, the Breslow--Peto estimator and its model-based and model-robust variance estimators are valid in the new
probability ratio model in both continuous and discrete time.

Both the odds and probability ratio mdoels can be generalized to regression settings, with multiple categorical or continuous covariates,
as discrete-time versions of Cox's proportional hazards models.
Further research is currently pursued to generalize the proposed methods and theory of estimation to such regression models.

For convenience, Table~\ref{tab:notation} lists the models and the point and variance estimators discussed in the remaing sections.

\begin{table}
\caption{Models and point and variance estimators} \label{tab:notation}  \vspace{-.02in}
\begin{center}
\begin{tabular*}{.8\textwidth}{@{\extracolsep\fill} cccc} \hline
      &                 &  Model-based       &  Model-robust  \\
Model & Point estimator & variance estimator & variance estimator \\ \hline
 \multicolumn{4}{c}{Analysis of $2\times 2$ tables (Section~\ref{sec:2by2})} \\
model (\ref{eq:oddr-model}) & $\hat \beta^{(w)}$ & $N_\bullet^{-1} \hat \Sigma^{(wb)}$ & $N_\bullet^{-1} \hat \Sigma^{(w)}$ \\
model (\ref{eq:pr-model})   & $\hat \gamma^{(w)}$  & $N_\bullet^{-1} \hat V^{(wb)}$ & $N_\bullet^{-1} \hat V^{(w)}$   \\
 \multicolumn{4}{c}{Two-sample survival analysis (Section~\ref{sec:survival})} \\
model (\ref{eq:oddr-model2}) & $\hat \beta^{(w)}$ & $N^{-1} \hat \Sigma^{(w2b)}$ & $N ^{-1} \hat \Sigma^{(w2)}$ \\
model (\ref{eq:pr-model2})   & $\hat \gamma^{(w)}$  & $N^{-1} \hat V^{(w2b)}$ & $N ^{-1} \hat V^{(w2)}$   \\ \hline
\end{tabular*}\\[1ex]
\parbox{.8\textwidth}{\scriptsize Note: $N_\bullet^{-1} \hat \Sigma^{(wb)} = N^{-1} \hat \Sigma^{(w2b)}$ and $N_\bullet^{-1} \hat V^{(wb)}=N^{-1} \hat V^{(w2b)}$ by definition.}
\end{center} \vspace{-.1in}
\end{table}

\section{Analysis of 2 by 2 tables} \label{sec:2by2}

Suppose that a series of $2 \times 2$ tables on a response and a factor are obtained independently from $J$ strata, as shown in Table~\ref{tab:2by2}.
Denote $N_j = N_{1j} + N_{2j}$, and $N_\bullet=\sum_{j=1}^J N_j$.
For concreteness, the value 1 is called a success, and 2 a failure for the response.
For each $j=1,\ldots,J$, the counts $n_{11j}$ and $n_{21j}$ are assumed to be independent binomial, with fixed denominators $N_{1j}$ and $N_{2j}$ and
unknown probabilities $p_{11j}$ and $p_{21j}$.
Denote $p_{12j} = 1-p_{11j}$ and $p_{22j} = 1-p_{21j}$. The raw estimates of probabilities are defined as
$\hat p_{11j} = n_{11j} / N_{1j}$, $\hat p_{21j} = n_{21j} / N_{2j}$, etc.

For asymptotic evaluation, it is of interest to examine two distinct settings, referred to as Settings I and II. In Setting I (large tables),
the number of tables $K$ is fixed while individual cell counts increase to infinity. In Setting II (many sparse tables),
the number of tables increases while the cell counts remain bounded.

\begin{table}
\caption{The $j$th $2 \times 2$ table} \label{tab:2by2}  \vspace{-.02in}
\begin{center}
\begin{tabular*}{.4\textwidth}{@{\extracolsep\fill} c ccc} \hline
& \multicolumn{2}{c}{response} \\
factor & 1 & 2  & total  \\ \hline
1      & $n_{11j}$ & $n_{12j}$ & $N_{1j}$ \\
2      & $n_{21j}$ & $n_{22j}$ & $N_{2j}$ \\  \hline
\end{tabular*}
\end{center}
\end{table}

\subsection{Odds ratio inference}

Consider a model on the odds ratios, $\psi_j^* = p_{11j} p_{22j} /(p_{12j} p_{21j})$, as follows (Zelen 1971; Breslow 1976):
\begin{align}
\log (\psi_j^*) = x_j^\T \beta^*, \quad j=1,\ldots,J, \label{eq:oddr-model}
\end{align}
where $x_j$ is a covariate vector associated with the $j$th table and $\beta^*$ is an unknown coefficient vector.
Inference in such models has been extensively studied, particularly in the case of common odds ratios,
$\psi_1^* = \cdots = \psi_J^* = \exp(\beta^*)$, corresponding to $x_1 =\cdots=x_J=1$ (e.g., Cochran 1954; Mantel \& Haenszel 1959).

For our method, we use the estimating function
\begin{align}
 \tau_\rho(\beta) = \sum_{j=1}^J \rho_j(\beta) \{ \hat p_{11j} \hat p_{22j} - \psi_j (\beta) \hat p_{12j} \hat p_{21j} \} x_j, \label{eq:EE}
\end{align}
where $\psi_j(\beta) = \exp(  x_j^\T \beta)$ and $\rho_j(\beta)$ is a scalar, non-random function of $\beta$ for $j=1,\ldots,J$.
This estimating function is apparently unbiased: $E\{ \tau_\rho(\beta) \}=0$ at $\beta=\beta^*$.
The associated estimator, denoted by $\hat\beta_\rho$, is defined as a solution to $\tau_\rho(\beta)=0$.
A choice of $\rho_j(\beta)$ independently of $\beta$ is $\rho_j^{(0)}(\beta) = N_{1j} N_{2j} / N_j$, which in the case of common odds ratios yields
the original Mantel--Haenszel estimator
$$
\hat \beta^{(0)} = \log \left(\frac{\sum_{j=1}^J n_{11j} n_{22j}  / N_j} {\sum_{j=1}^J n_{12j} n_{21j}  / N_j} \right).
$$
In general, an Mantel--Haenszel estimator, $\hat\beta^{(0)}$, can be defined as a solution to (\ref{eq:EE}) for $\rho_j=\rho_j^{(0)}$, with possibly stratum-dependent covariates.

We develop our method in several steps. First, we find the optimal choice of $\rho_j(\beta)$ in minimizing the asymptotic variance of $\hat \beta_\rho$ in Setting I, provided that
model (\ref{eq:oddr-model}) is valid. Then we derive a simple choice of $\rho_j(\beta)$, defined as
\begin{align}
\rho^{(w)}_j(\beta) = \frac{N_{1j} N_{2j}} { N_{1j} \psi_j (\beta)  + N_{2j}}, \label{eq:rho}
\end{align}
such that it not only provides a reasonable approximation to the optimal choice but also leads to a desirable reduction of (\ref{eq:EE})
to conditional likelihood estimation (Breslow 1981; McCullagh \& Nelder 1989) when the total number of successes happens to be 1, $n_{11j}+n_{21j}=1$.
In general, conditional likelihood estimation is well-behaved in Setting II,
and closely related to partial likelihood estimation (Cox 1972) in survival analysis, which is discussed in Section~\ref{sec:survival}.
The resulting estimator, $\hat\beta^{(w)}$, defined as a solution to  $\tau^{(w)}(\beta)=0$ is called the
weighted Mantel--Haenszel estimator, where
\begin{align}
 \tau^{(w)}(\beta) = \sum_{j=1}^J \frac{n_{11j}n_{22j} - \psi_j(\beta) n_{12j}n_{21j}} { N_{1j} \psi_j (\beta)  + N_{2j}} x_j. \label{eq:WEE}
\end{align}
Incidentally, $\hat\beta^{(w)}$ can be directly shown to be  a maximizer of the concave function
\begin{align}
\sum_{j=1}^J \big[ N_{1j} \hat A_j x_j^\T \beta - (N_{1j} \hat A_j + N_{2j} \hat B_j) \log \{ N_{1j} \psi_j(\beta) + N_{2j}\} \big], \label{eq:WEE-loss}
\end{align}
where $\hat A_j = \hat p_{11j} \hat p_{22j}$ and $\hat B_j = \hat p_{12j} \hat p_{21j}$.
Finally, to complement model-based inference, we propose a model-robust estimator of the variance of $\hat\beta^{(w)}$, which is consistent in both Settings I and II while allowing
for possible misspecification of model (\ref{eq:oddr-model}).

\begin{rem} \label{rem:est-invariance}
The estimator $\hat \beta^{(0)}$ is invariant to the exchange between the factor levels and between the response values with constant covariates.
But such invariance in general fails with nonconstant covariates.
By comparison, the weighted Mantel--Haenszel  estimator $\hat \beta^{(w)}$ is invariant to the exchange between the factor levels with possibly nonconstant covariates,
but generally not to the exchange between the response values (except, for example, $N_{1j}=N_{2j}$ for all $j$).
In fact, it is preferable to apply the estimator $\hat \beta^{(w)}$ with relatively small success probabilities $(p_{11j}, p_{21j})$, as discussed later in Section~\ref{sec:oddr-point}.
\end{rem}

\subsubsection{Point estimation} \label{sec:oddr-point}

In this section, we discuss the derivation of the choice $\rho^{(w)}_j(\beta)$ for the weighted Mantel--Haenszel estimator $\hat\beta^{(w)}$.
First, it can be shown  that if model (\ref{eq:oddr-model}) is valid, then
the asymptotic variance of $\hat\beta_\rho$ is of the sandwich form  $N_\bullet^{-1} H_\rho^{-1} (\beta^*) G_\rho (\beta^*)H_\rho^{-1}(\beta^*)$ under standard regularity
conditions in both Settings I and II (Davis 1985), where
\begin{align*}
G_\rho(\beta) & = N_\bullet^{-1}\sum_{j=1}^J \rho_j^2(\beta) \var \{ \hat p_{11j} \hat p_{22j} - \psi_j (\beta) \hat p_{12j} \hat p_{21j} \} x_j x_j^\T,\\
H_\rho(\beta) & = N_\bullet^{-1}\sum_{j=1}^J \rho_j(\beta)  \psi_j(\beta) p_{12j} p_{21j}  x_j x_j^\T.
\end{align*}
Then the optimal choice of $\rho_j(\beta)$ can be obtained similarly as in theory of quasi-likelihood functions (McCullagh \& Nelder 1989). See the Supplement for a direct proof.

\begin{pro} \label{pro:optimal-EE}
Suppose that odds ratio model (\ref{eq:oddr-model}) is valid. \\
(i) The asymptotic variance  of $\hat\beta_\rho$ in both Settings I and II is minimized by the choice
\begin{align*}
\rho_j^\dag (\beta) = \frac{\psi_j(\beta) p_{12j} p_{21j} }{ \var \{ \hat p_{11j} \hat p_{22j} - \psi_j (\beta) \hat p_{12j} \hat p_{21j} \}}, \quad j=1,\ldots,J.
\end{align*}
(ii) In Setting I as $ N_j\to \infty$ and $N_{j1} /N_j$ tending to a constant in $(0,1)$ for each $j$ with $J$ fixed, the asymptotic variance  of $\hat\beta_\rho$
is also minimized by the choice
\begin{align*}
\rho_j^\ddag (\beta) = \frac{N_{1j} N_{2j} }{ N_{1j} \{ p_{11j} + \psi_j(\beta) p_{12j}\} + N_{2j} \{\psi_j(\beta) p_{21j} +  p_{22j}\} }, \quad j=1,\ldots,J.
\end{align*}
\end{pro}

\vspace{.05in}
The foregoing choice $\rho_j^\ddag(\beta)$ cannot be directly used, due to its dependency on the unknown quantities  $(p_{11j},p_{21j})$.
In Setting I, this difficulty can in principle be overcome by replacing $(p_{11j},p_{21j})$ with their consistent estimators
$(\hat p_{11j}, \hat p_{21j})$.
The resulting estimator of $\beta$ can be shown to achieve the same asymptotic variance as the (infeasible) estimator $\hat\beta_{\rho^\ddag}$ with the optimal choice $\rho^\ddag_j(\beta)$.
In Setting II, however, such data-dependent approximation of $\rho_j^\ddag (\beta)$ can lead to poor performance,
because the variation of $(\hat p_{11j}, \hat p_{21j})$ is no longer negligible with bounded sizes $(N_{1j},N_{2j})$.

To achieve good performance in both Settings I and II, we propose the simple choice $\rho_j^{(w)}(\beta)$, defined in (\ref{eq:rho}),
as a data-independent (i.e., non-random) approximation to $\rho_j^\ddag( \beta)$.
The relative error in this approximation for $\beta = \beta^*$,
\begin{align*}
\frac{ \rho_j^{(w)}(\beta^*)} {\rho_j^\ddag(\beta^*) } -1 = \{1- \psi_j(\beta^*)\}\frac{N_{1j} p_{11j} - N_{2j}p_{21j} }{N_{1j} \psi_j (\beta^*)  + N_{2j}},
\end{align*}
is close to 0 whenever the odds ratio $\psi_j(\beta^*)$ is close to 1 or the success probabilities $(p_{11j}, p_{21j})$ are close to 0.
The resulting estimator $\hat \beta^{(w)}$ is expected to perform similarly to the (infeasible) optimal estimator $\hat\beta_{\rho^\ddag}$ in Setting I, especially
when differences between the two groups are small or positive responses are rare.

The appropriateness of the proposed choice $\rho^{(w)}_j(\beta)$ in Setting II (as well as Setting I) can also be seen from the following connection to conditional likelihood estimation,
which is known for its superior performance in both Settings I and II (Breslow 1981; McCullagh \& Nelder 1989). In fact, the condition score function is
\begin{align}
s(\beta) = \sum_{j=1}^J \left\{ n_{11j} - \mu_{11j}(\beta) \right\} x_j, \label{eq:cond-score}
\end{align}
where $\mu_{11j}(\beta)$ is the conditional expectation of $n_{11j}$ given the marginals $(n_{11j}+n_{21j}, N_{1j},N_{2j})$ in the $j$th table with odds ratio $\psi_j(\beta)$.
In the case of $n_{11j}+n_{21j}=1$ (i.e., a total of one success in the two groups), it can be directly shown that
\begin{align}
 & n_{11j} - \mu_{11j}(\beta) =   n_{11j} - \frac{(n_{11j}+n_{21j})N_{1j} \psi_j(\beta)}{ N_{1j} \psi_j(\beta) + N_{2j} } \nonumber  \\
 & =  \frac{n_{11j}n_{22j} - \psi_j(\beta) n_{12j}n_{21j} + \{1-\psi_j(\beta)\} n_{11j} n_{21j}} { N_{1j} \psi_j (\beta)  + N_{2j}} \nonumber \\
 & =  \frac{n_{11j}n_{22j} - \psi_j(\beta) n_{12j}n_{21j} } { N_{1j} \psi_j (\beta)  + N_{2j}} ,  \label{eq:one-success}
\end{align}
where the last step holds because $(n_{11j},n_{21j}) = (0,1)$ or $(1,0)$ and hence $n_{11j} n_{21j}=0$.
Therefore, the $j$th contribution to the proposed estimating function $\tau^{(w)}(\beta)$ in (\ref{eq:WEE}) coincides with
that to the conditional score function $s(\beta)$ when the total number of successes is 1 in the $j$th table.
The coincidence in such an extreme case suggests that the proposed estimator $\hat\beta^{(w)}$ tends to perform similarly to the maximum
conditional likelihood estimator $\hat \beta^{(c)}$, defined as a solution to $s(\beta)=0$.

There is another implication from the discussion above on
the approximation of $\hat\beta^{(w)}$ to $\hat\beta_{\rho^\ddag}$ with small $(p_{11j},p_{21j})$ in Setting II
and the reduction to $\hat\beta^{(c)}$ in the case of a total of one success.
It is more desirable to apply $\hat\beta^{(w)}$ with relatively small success probabilities for responses. Otherwise,
the labeling of response values should be exchanged.

\begin{rem} \label{rem:CML}
There are several reasons why the proposed estimator $\hat\beta^{(w)}$ can be worthwhile compared with the
conditional likelihood estimator $\hat \beta^{(c)}$, even though $\hat\beta^{(c)}$ seems attractive on various grounds (e.g., Lindsay 1980; Hauck 1984).
First, it is straightforward to compute $\hat\beta^{(w)}$ with estimating function  $\tau^{(w)}(\beta)$ in a closed form,
whereas computation of $\hat\beta^{(c)}$ is difficult for tables with large marginals due to the complexity in numerical evaluation of $\mu_{11j}(\beta)$.
See Breslow \& Cologne (1986) and McCullagh \& Nelder (1989) for approximate methods for $\hat\beta^{(c)}$.
Second and more importantly, $\tau^{(w)}(\beta)$ is an unbiased estimating function under model (\ref{eq:oddr-model}),
and hence consistency can be directly established by standard large sample theory.
The approximate methods for $\hat\beta^{(c)}$ are numerically motivated, but their statistical properties remain to be studied.
Finally, analytical simplicity of $\tau^{(w)}(\beta)$ also facilitates model-robust variance estimation
with possible misspecification of model (\ref{eq:oddr-model}) as discussed in Section \ref{sec:oddr-var}.
\end{rem}

\begin{rem} \label{rem:davis}
Davis (1985) considered a class of estimating functions,
\begin{align}
\sum_{j=1}^J  \{ g_{1j}(\beta) \hat p_{11j} \hat p_{22j} - g_{2j}(\beta) \psi_j (\beta) \hat p_{12j} \hat p_{21j} \} x_j,  \label{eq:EE-davis}
\end{align}
where $g_{1j}(\beta)= g(\beta;n_{11j}, n_{12j}, n_{21j}, n_{22j}) $ and $g_{2j}(\beta) = g(\beta;n_{11j}+1, n_{12j}-1, n_{21j}-1, n_{22j}+1) $ for some
scalar, possibly data-dependent function $g$. It is shown that (\ref{eq:EE-davis}) is an unbiased estimating function, conditionally on the marginals $(n_{11j}+n_{21j}, N_{1j},N_{2j})$, $j=1,\ldots,J$,
although such conditional unbiasedness seems not further pursued.
In fact, an optimal choice of $g$, restricted to be non-random and hence $g_{1j}=g_{2j}$, is stated, without proof, first in the same form as $\rho^\dag_j(\beta)$ in Proposition~\ref{pro:optimal-EE},
and then in an expression which appears to disagree with our calculation in the proof of Proposition~\ref{pro:var-est}(ii).
Nevertheless, a concrete choice of $g$ was then proposed, but differently from our choice $\rho_j^{(w)}$.
In numerical examples of Breslow \& Cologne (1986), the estimator of Davis (1985) was found to sometimes differ noticeably from the conditional likelihood estimator $\hat\beta^{(c)}$.
\end{rem}

\subsubsection{Variance estimation} \label{sec:oddr-var}

In this section, we propose a consistent estimator of the asymptotic variance of the weighted Mantel--Haenszel estimator $\hat\beta^{(w)}$ in both Settings I and II, while allowing
for possible misspecification of model (\ref{eq:oddr-model}).
Such a variance estimator for a point estimator is referred to as model-robust with respect to the associated model.
In contrast, model-based variance estimators are constructed such that they are inconsistent unless the associated model is correctly specified.
There are various model-based variance estimators for the Mantel--Haenszel estimator $\hat\beta^{(0)}$ under the assumption of common odds ratios.
See Huritz et al.~(1988) for a review and the Supplement for further discussion about relationships between existing variance estimators.

To describe the asymptotic behavior of $\hat\beta^{(w)}$, we adopt the standard theory of estimation in misspecified models (e.g., White 1982; Manski 1988).
First, it can be shown in both Settings I and II, under regularity conditions similar to those in Davis (1985), that
$\hat \beta^{(w)}$ converges in probability to a target value $\bar\beta^{(w)}$, defined as a solution to
\begin{align}
 0 = \sum_{j=1}^J \rho_j^{(w)}(\beta) \{ p_{11j} p_{22j} - \psi_j (\beta) p_{12j}  p_{21j} \} x_j , \label{eq:WEE-pop}
\end{align}
or equivalently as a unique maximizer of the concave function
\begin{align}
\sum_{j=1}^J \big[ N_{1j} A_j x_j^\T \beta - (N_{1j} A_j + N_{2j} B_j) \log \{ N_{1j} \psi_j(\beta) + N_{2j}\} \big],  \label{eq:WEE-loss-pop}
\end{align}
where $ A_j =  p_{11j} p_{22j}$ and $ B_j =  p_{12j} p_{21j}$. Equation (\ref{eq:WEE-pop}) and function (\ref{eq:WEE-loss-pop})
are the population versions of (\ref{eq:WEE}) and (\ref{eq:WEE-loss}) respectively.
If model (\ref{eq:oddr-model}) is valid, then $\bar\beta^{(w)} = \beta^*$ such that $\psi_j(\beta^*) = \psi_j^*$ for $j=1,\ldots,J$.
Otherwise, $\psi_j (\bar \beta^{(w)})$ may differ from $\psi_j^*$.
Moreover, it can be shown that
$ N_\bullet^{-1/2} (\hat \beta^{(w)} - \bar \beta^{(w)})$ converges in distribution to $\N(0, \Sigma^{(w)})$, where
$\Sigma^{(w)} =  H^{(w)-1}(\beta) G^{(w)}(\beta) H^{(w)-1}(\beta) |_{\beta= \bar \beta^{(w)}}$ with
\begin{align*}
G^{(w)} (\beta) & = N_\bullet^{-1} \sum_{j=1}^J \rho_j^{(w)2}(\beta) \var \{ \hat p_{11j} \hat p_{22j} - \psi_j (\beta) \hat p_{12j} \hat p_{21j} \} x_j x_j^\T,\\
H^{(w)}  (\beta) & = N_\bullet^{-1} \sum_{j=1}^J  (N_{1j} A_j + N_{2j} B_j)\frac{N_{1j}N_{2j}  \psi_j(\beta)}{ \{N_{1j} \psi_j(\beta) + N_{2j}\}^2} x_j x_j^\T .
\end{align*}
The matrix $H^{(w)}(\beta)$ is obtained from the negative Hessian of function (\ref{eq:WEE-loss-pop}). The asymptotic variance of $\hat \beta^{(w)}$
is $N_\bullet^{-1}  \Sigma^{(w)} =N_\bullet^{-1} H^{(w)-1}(\beta) G^{(w)}(\beta) H^{(w)-1}(\beta) |_{\beta= \bar \beta^{(w)}} $,
which is invariant if scaling by $N_\bullet^{-1}$ is dropped from the right hand side and from $G^{(w)}$ and $H^{(w)}$.

For the variance matrix $\Sigma^{(w)}$, our proposed estimator is
\begin{align*}
\hat \Sigma^{(w)} = \hat H^{(w)-1}(\beta) \hat G^{(w)}(\beta) \hat H^{(w)-1}(\beta) |_{\beta= \hat \beta^{(w)}},
\end{align*}
where
$\hat H^{(w)}(\beta) $ is defined as  $H^{(w)}(\beta)$  with $(A_j,B_j)$ replaced by $(\hat A_j,\hat B_j)$, but
$\hat G^{(w)} (\beta) = \sum_{j=1}^J \rho_j^{(w)2}(\beta) \hat \sigma_j(\beta) x_j x_j^\T$ with
\begin{align*}
\hat \sigma_j(\beta) & = \frac{ \hat p_{11j}\hat p_{12j}}{N_{1j}-1} \{\hat p_{22j} + \psi_j(\beta) \hat p_{21j} \}^2 + \frac{ \hat p_{21j} \hat p_{22j}}{N_{2j}-1} \{\hat p_{11j} + \psi_j(\beta) \hat p_{12j} \}^2 \\
& \quad - \{\psi_j(\beta)-1\}^2 \frac{ \hat p_{11j}\hat p_{12j}}{N_{1j}-1} \frac{ \hat p_{21j} \hat p_{22j}}{N_{2j}-1} , \quad j=1,\ldots, J.
\end{align*}
Then the following properties can be established.

\begin{pro} \label{pro:var-est}
Let $\sigma_j(\beta) = \var \{ \hat p_{11j} \hat p_{22j} - \psi_j (\beta) \hat p_{12j} \hat p_{21j} \}$.
 Assume that $N_{1j} \ge 2$ and $N_{2j} \ge 2$ for each $j=1,\ldots,J$. \\
(i) $\hat \sigma_j(\beta) \ge 0$ for any fixed $\beta$ and $j=1,\ldots,J$. \\
(ii) $ \hat \sigma_j(\beta)$ is an unbiased estimator of $\sigma_j(\beta)$ for any fixed $\beta$ and $j=1,\ldots,J$.\\
(iii) $\hat \Sigma^{(w)}$ is a consistent estimator of $\Sigma^{(w)}$ in both Settings I and II, with possible misspecification of model (\ref{eq:oddr-model}).
\end{pro}

The estimator $\hat \sigma_j(\beta)$ serves as a finite-sample correction to the simpler version
\begin{align}
\tilde\sigma_j(\beta) = \frac{ \hat p_{11j}\hat p_{12j}}{N_{1j}} \{\hat p_{22j} + \psi_j(\beta) \hat p_{21j} \}^2 + \frac{ \hat p_{21j} \hat p_{22j}}{N_{2j}} \{\hat p_{11j} + \psi_j(\beta) \hat p_{12j} \}^2 .
\end{align}
The estimator $\hat \Sigma^{(w)}$  with $ \hat \sigma_j(\beta)$ replaced by $\tilde\sigma_j(\beta)$ remains a consistent estimator of $\Sigma^{(w)}$ in Setting I,
but in general becomes inconsistent in Setting II with bounded $(N_{1j}, N_{2j})$.
Similar formulas to $\tilde\sigma_j(\beta)$ above can be found in Guilbaud (1983).

\begin{rem} \label{rem:robins-var}
It is instructive to compare the  variance estimator $\hat \Sigma^{(w)}$ with model-based variance estimators.
For the Mantel--Haenszel estimator $\hat\beta^{(0)}$, Robins et al.~(1986) proposed a variance estimator which is consistent in both Settings I and II
under the assumption of common odds ratios.
Their estimator of $\sigma_j(\beta)$ is defined as
\begin{align*}
\hat \sigma^{(b)}_j(\beta) & =  \frac{\psi_j(\beta) \hat p_{12j}\hat p_{21j}}{N_{1j}} \{\hat p_{22j} + \psi_j(\beta) \hat p_{21j} \}
 +  \frac{\hat p_{11j} \hat p_{22j}}{ N_{2j}} \{\hat p_{11j} + \psi_j(\beta) \hat p_{12j} \} .
\end{align*}
With possibly nonconstant covariates, it can be shown that if model (\ref{eq:oddr-model}) is valid, then a consistent estimator of $\Sigma^{(w)}$ in both Settings I and II is
\begin{align*}
\hat \Sigma^{(wb)} = \hat H^{(w)-1}(\beta) \hat G^{(wb)}(\beta) \hat H^{(w)-1}(\beta) |_{\beta= \hat \beta^{(w)}},
\end{align*}
where $\hat H^{(w)}(\beta) $ is as before, but
$\hat G^{(wb)} (\beta) = N_\bullet^{-1} \sum_{j=1}^J \rho_j^{(w)2}(\beta) \hat \sigma^{(b)}_j(\beta) x_j x_j^\T$. Hence
$\hat \Sigma^{(w)}$ would be identical to $\hat \Sigma^{(wb)}$, except that $\hat \sigma_j(\beta)$ is in place of $\hat \sigma^{(b)}_j(\beta)$.
In fact, $\hat \sigma_j(\beta)$ is unbiased for $\sigma_j(\beta)$ with any fixed $\beta$ as shown in Proposition~\ref{pro:var-est},
whereas $\hat \sigma^{(b)}_j (\beta)$ is unbiased for $\sigma_j(\beta)$ only with $\beta= \beta^*$ such that $\psi_j (\beta^*) = \psi^*_j$.
Algebraically,  $\hat \sigma_j(\beta)$ is a bivariate polynomial of $(\hat p_{11j}, \hat p_{21j})$ including a term $\hat p_{11j}^2\hat p_{21j}^2$ of total degrees 4,
whereas  $\hat \sigma^{(b)}_j(\beta)$ involves terms only up to 3 total degrees such as $\hat p_{11j}^2\hat p_{21j}$ and $\hat p_{11j}\hat p_{21j}^2$.
\end{rem}

\begin{rem} \label{rem:var-invariance}
Both the variance estimators  $\hat \Sigma^{(w)}$ and $ \hat \Sigma^{(wb)}$ are invariant to the exchange between the factor levels, but generally not to that between
the response values. This is similar to the invariance properties of the point estimator $\hat\beta^{(w)}$, discussed in Remark \ref{rem:est-invariance}.
\end{rem}

\begin{rem} \label{rem:orginal-MH}
For the Mantel--Haenszel estimator $\hat\beta^{(0)}$, similar results can be obtained. In fact,
a model-robust estimator for the asymptotic variance of $\hat\beta^{(0)}$ is
 $\hat \Sigma (\hat\beta^{(0)}) =\hat H^{(0)-1}(\beta)$ $\times \hat G^{(0)}(\beta) \hat H^{(0)-1}(\beta) |_{\beta= \hat \beta^{(0)}} $,
and a model-based estimator is
 $\hat \Sigma^{(b)} (\hat\beta^{(0)}) =\hat H^{(0)-1}(\beta) \hat G^{(0b)}(\beta) $ $ \times \hat H^{(0)-1}(\beta) |_{\beta= \hat \beta^{(0)}} $,
 where $\hat H^{(0)}(\beta) = \sum_{j=1}^J \rho_j^{(0)}(\beta)\psi_j(\beta) \hat p_{12j} \hat p_{21j}  x_j x_j^\T$,
$\hat G^{(0)} (\beta) = \sum_{j=1}^J $ $\rho_j^{(0)2}(\beta) \hat \sigma_j(\beta) x_j x_j^\T$,
and $\hat G^{(0b)} (\beta) = $ $\sum_{j=1}^J \rho_j^{(0)2}(\beta) \hat \sigma^{(b)}_j(\beta) x_j x_j^\T$.
With nonconstant covariates, these variance estimators are generally not invariant to either the exchange between the factor levels or
between the response values, similarly as the point estimator $\hat\beta^{(0)}$ discussed in Remark~\ref{rem:est-invariance}.
With constant covariates, however, $\hat \Sigma (\hat\beta^{(0)})$ becomes invariant to the exchange between the factor levels and between the response values,
similarly as the point estimator $\hat\beta^{(0)}$. The estimator $\hat \Sigma^{(b)} (\hat\beta^{(0)})$, which reduces to a variance estimator proposed in Robins et al.~(1986),
remains invariant to the exchange between the factor levels, but not to that between
the response values. Hence a symmetrized version of $\hat \Sigma^{(b)} (\hat\beta^{(0)})$ is also proposed in Robins et al.~(1986) to achieve two-way invariance.
\end{rem}

\subsection{Probability ratio inference}

Consider a model on the probability ratios, $\phi_j^* = p_{11j} /p_{21j}$, as follows:
\begin{align}
\log (\phi_j^*) = x_j^\T \gamma^*, \quad j=1,\ldots,J, \label{eq:pr-model}
\end{align}
where $x_j$ is a covariate vector associated with the $j$th table and $\gamma^*$ is an unknown coefficient vector.
Compared with odds ratio model (\ref{eq:oddr-model}), such probability ratio models have been directly studied to a lesser extent for various reasons.
First, odds ratios models are popular, especially in retrospective studies, due to the invariance of odds ratios to prospective or retrospective sampling.
Second, the availability of conditional likelihood inference given table marginals can be appealing in model (\ref{eq:oddr-model}), regarding elimination of nuisance parameters.
Third, odds ratios are often considered an approximation to probability ratios when success probabilities are small, corresponding to rare diseases in biomedical applications.
Nevertheless, odds ratios are persistently biased estimates of probability ratios in being further away from 1, unless the probabilities are identical between the two factor levels.
Moreover, as discussed below, model-robust inference in model (\ref{eq:pr-model}) can be carried with carefully constructed estimating functions, in a parallel manner to that in model (\ref{eq:oddr-model}).
There is also a remarkable connection to Breslow--Peto modification of partial likelihood estimation with tied event times.

\vspace{-.05in}
\subsubsection{Point estimation} \label{sec:pr-point}

\vspace{-.1in}
We use the estimating function \vspace{-.in}
\begin{align}
 \zeta_q(\gamma) = \sum_{j=1}^J q_j(\gamma) \{ \hat p_{11j} - \phi_j (\gamma) \hat p_{21j} \} x_j, \label{eq:pr-EE}
\end{align}
where $\phi_j(\gamma) = \exp(  x_j^\T \gamma)$ and $q_j(\gamma)$ is a scalar, non-random function of $\gamma$ for $j=1,\ldots,J$.
This estimating function is unbiased: $E\{ \zeta_q(\gamma) \}=0$ at $\gamma=\gamma^*$.
The associated estimator, denoted by $\hat\gamma_q$, is defined as a solution to $\zeta_q(\gamma)=0$.
A choice of $q_j(\gamma)$ similar to $\rho^{(0)}_j$ is $q_j^{(0)}(\gamma) = N_{1j} N_{2j} / N_j$, independently of $\gamma$,
which in the case of common probability ratios yields  $\hat\gamma^{(0)} = \log\{(\sum_{j=1}^J  n_{11}N_{2j} /N_j)/ ( \sum_{j=1}^J n_{21j}N_{1j} /N_j)\}$.

Our proposed choice of $q_j(\gamma)$ is similar to $ \rho^{(w)}_j$: \vspace{-.05in}
\begin{align}
q^{(w)}_j(\gamma) = \frac{N_{1j} N_{2j}} { N_{1j} \phi_j (\gamma)  + N_{2j}}. \label{eq:qfcn}
\end{align}
The resulting estimator, $\gamma^{(w)}$, is defined as a solution to $\zeta^{(w)}(\gamma)=0$, where
\begin{align}
 \zeta^{(w)}(\gamma) = \sum_{j=1}^J \frac{n_{11j} N_{2j} - \phi_j(\gamma) n_{21j} N_{1j} } { N_{1j} \phi_j (\gamma)  + N_{2j}} x_j, \label{eq:pr-WEE}
\end{align}
which can be equivalently rewritten as\vspace{-.1in}
\begin{align}
\zeta^{(w)}(\gamma) = \sum_{j=1}^J \left\{ n_{11j} - \frac{(n_{11j}+n_{21j}) N_{1j} \phi_j (\gamma)  } { N_{1j} \phi_j (\gamma)  + N_{2j}} \right\} x_j,\label{eq:pr-WEE2}
\end{align}
Moreover, $\hat\gamma^{(w)}$ can be directly shown to be a maximizer of the concave function
\begin{align}
\sum_{j=1}^J \big[ n_{11j}  x_j^\T \gamma - (n_{11j}+n_{21j}) \log \{ N_{1j} \phi_j(\gamma) + N_{2j}\} \big], \label{eq:pr-WEE-loss}
\end{align}
which is the log-likelihood of a pseudo-model that the $n_{11j} +n_{21j}$ successes are independent and identically distributed Bernoulli, each with probability
$ N_{1j} \phi_j(\gamma) /\{ N_{1j} \phi_j(\gamma) + N_{2j}\} $ from factor level 1 and the remaining probability from factor level 2.

The derivation of our choice $q^{(w)}_j(\gamma)$ can be seen from two angles, similarly as in Section~\ref{sec:oddr-point}.
One is based on an approximation to the optimal choice of $q_j(w)$ in Setting I with model (\ref{eq:pr-model}) correctly specified.
In fact, it can be shown that if model (\ref{eq:pr-model}) is valid, then
the asymptotic variance of $\hat\gamma_q$ is of the sandwich form  $N_\bullet^{-1}  D_q^{-1} (\gamma^*) C_q (\gamma^*)D_q^{-1}(\gamma^*)$ under standard regularity
conditions in both Settings I and II, where
\begin{align*}
C_q(\gamma) & = N_\bullet^{-1}  \sum_{j=1}^J q_j^2(\gamma) \var \{ \hat p_{11j}  - \phi_j (\gamma)  \hat p_{21j} \} x_j x_j^\T,\\
D_q(\gamma) & = N_\bullet^{-1} \sum_{j=1}^J q_j(\gamma)  \phi_j(\gamma)  p_{21j}  x_j x_j^\T.
\end{align*}
The optimal choice of $q_j(\gamma)$ can be obtained as follows, similarly as in Proposition~\ref{pro:pr-optimal-EE}.
There is however a subtle difference: $q_j^\ddag (\gamma)$ is optimal in both Settings I and II with $q_j^\ddag (\gamma^*) =q_j^\dag (\gamma^*)$ exactly,
whereas $\rho_j^\ddag(\beta)$ is optimal in Setting I but not Setting II.

\begin{pro} \label{pro:pr-optimal-EE}
Suppose that odds ratio model (\ref{eq:oddr-model}) is valid.  The asymptotic variance  of $\hat\gamma_q$ in both Settings I and II is minimized by the choice
\begin{align*}
q_j^\dag (\gamma) &= \frac{\phi_j(\gamma) p_{21j} }{ \var \{ \hat p_{11j} - \phi_j (\gamma) \hat p_{21j} \}}, \quad j=1,\ldots,J,
\end{align*}
or equivalently by the choice
\begin{align*}
q_j^\ddag (\gamma) = \frac{N_{1j} N_{2j} }{ N_{1j}\phi_j (\gamma)  p_{22j}  + N_{2j}p_{12j}  }, \quad j=1,\ldots,J.
\end{align*}
\end{pro}

By Proposition~\ref{pro:pr-optimal-EE}, the relative error of the data-independent choice $q_j^{(w)}(\gamma)$ as an approximation to the optimal choice $q_j^\ddag(\gamma)$ for $\gamma=\gamma^*$ is
\begin{align*}
\frac{ q_j^{(w)}(\gamma^*)} {q_j^\ddag(\gamma^*) } -1 = - \frac{N_{1j}\phi_j (\gamma^*)  p_{21j} + N_{2j}p_{11j} }{N_{1j} \phi_j (\gamma^*)  + N_{2j}},
\end{align*}
which is close to 0 whenever the success probabilities $(p_{11j}, p_{21j})$ are close to 0.
Hence the proposed estimator $\hat\gamma^{(w)}$ is expected to perform close to being optimal in both Settings I and II, especially when positive responses are rare.

The second motivation for our choice $q_j^{(w)}(\gamma)$ is that the resulting estimator $\hat\gamma^{(w)}$ coincides with the
maximum partial likelihood estimator with Breslow's (1974) and Peto's (1972) modification for tied death times in two-sample survival analysis.
The $j$th table can be constructed from the death and survival counts by the factor levels among the risk set at the $j$th death time.
The estimating function (\ref{eq:pr-WEE2}) or the criterion function (\ref{eq:pr-WEE-loss}) is identical
to the score function or the log-likelihood function in the Breslow--Peto modification of partial likelihood estimation, as shown in Proposition~\ref{pro:breslow-peto}.
Henceforth, the estimator $\hat\gamma^{(w)}$ can be referred to as the Breslow--Peto estimator.

The preceding coincidence need to be carefully understood.
In the case of a total of one success in $j$th table, the $j$th contribution of $\tau^{(w)}(\beta)$
and that of $\zeta^{(w)}(\gamma)$ are both identical to that of the conditional score function $s(\beta)$.
In general, with total numbers of successes greater than 1, the three functions $s(\beta)$, $\tau^{(w)}(\beta)$, and $\zeta^{(w)}(\gamma)$ differ from each other.
The first two $s(\beta)$ and $\tau^{(w)}(\beta)$ lead to consistent estimation of $\beta^*$ under odds ratio model (\ref{eq:oddr-model}).
In contrast, the estimating function  $\zeta^{(w)}(\gamma)$ leads to consistent estimation of $\gamma^*$ under probability ratio model (\ref{eq:pr-model}),
even though it was considered to approximate the conditional score $s(\beta)$ in the context of Cox's (1972) discrete-time proportional hazards model.
See Remark \ref{rem:oddr-var-special} and Section~\ref{sec:survival} for further discussion.

\subsubsection{Variance estimation} \label{sec:pr-var}

We present two estimators of the asymptotic variance of the Breslow--Peto estimator $\hat\gamma^{(w)}$ in both Settings I and II.
One is a model-based variance estimator, consistent provided that model (\ref{eq:pr-model}) is valid.
The other is a model-robust variance estimator, consistent in the presence of possible model misspecification.

First, we describe the asymptotic behavior of $\hat\gamma^{(w)}$ while allowing for possible misspecification of model (\ref{eq:pr-model}).
By standard theory of estimation with model misspecification (e.g., White 1982; Manski 1988), it can be shown in both Settings I and II
 that $\hat \gamma^{(w)}$ converges in probability to a target value $\bar\gamma^{(w)}$, defined as a solution to
\begin{align}
 0 = \sum_{j=1}^J q_j^{(w)}(\gamma) \{ p_{11j}  - \phi_j (\gamma)  p_{21j} \} x_j , \label{eq:pr-WEE-pop}
\end{align}
or equivalently as a unique maximizer of the concave function
\begin{align}
\sum_{j=1}^J \big[ N_{1j} p_{11j} x_j^\T \gamma - (N_{1j} p_{11j} + N_{2j} p_{21j}) \log \{ N_{1j} \phi_j(\gamma) + N_{2j}\} \big].  \label{eq:pr-WEE-loss-pop}
\end{align}
Equation (\ref{eq:pr-WEE-pop}) and function (\ref{eq:pr-WEE-loss-pop})
are the population versions of (\ref{eq:pr-WEE}) and (\ref{eq:pr-WEE-loss}) respectively.
If model (\ref{eq:pr-model}) is valid, then $\bar\gamma^{(w)} = \gamma^*$ such that $\phi_j(\gamma^*) = \phi_j^*$ for $j=1,\ldots,J$. Otherwise, $\phi_j (\bar \gamma^{(w)})$ may differ from $\phi_j^*$.
Moreover, it can be shown that
$ N_\bullet^{-1/2} (\hat \gamma^{(w)} - \bar \gamma^{(w)})$ converges in distribution to $\N(0,  V^{(w)})$, where
$ V^{(w)} = D^{(w)-1}(\gamma) C^{(w)}(\gamma) D^{(w)-1}(\gamma) |_{\gamma= \bar \gamma^{(w)}}$ with
\begin{align*}
C^{(w)} (\gamma) & = N_\bullet^{-1} \sum_{j=1}^J  q_j^{(w)2}(\gamma) \var \{ \hat p_{11j}  - \phi_j (\gamma)  \hat p_{21j} \} x_j x_j^\T,\\
D^{(w)}  (\gamma) & = N_\bullet^{-1} \sum_{j=1}^J  (N_{1j} p_{11j} + N_{2j} p_{21j} )\frac{N_{1j}N_{2j}  \phi_j(\gamma)}{ \{N_{1j} \phi_j(\gamma) + N_{2j}\}^2} x_j x_j^\T .
\end{align*}
The matrix $D^{(w)}(\gamma)$ is obtained from  the negative Hessian of function (\ref{eq:pr-WEE-loss-pop}). The asymptotic variance of $\hat \gamma^{(w)}$
is then $N_\bullet^{-1}  V^{(w)} =N_\bullet^{-1} D^{(w)-1}(\gamma) C^{(w)}(\gamma) D^{(w)-1}(\gamma) |_{\gamma= \bar \gamma^{(w)}} $.

For the variance matrix $ V^{(w)}$, our model-robust estimator is
$\hat  V^{(w)} =  \hat D^{(w)-1}(\gamma) \hat C^{(w)}(\gamma) $ $\times \hat D^{(w)-1}(\gamma) |_{\gamma= \hat \gamma^{(w)}}$,
where
$\hat D^{(w)}(\gamma) $ is defined as  $D^{(w)}(\gamma)$  with $(p_{11j},p_{21j})$ replaced by $(\hat p_{11j},\hat p_{21j})$, but
$\hat C^{(w)} (\gamma) = N_\bullet^{-1}  \sum_{j=1}^J  q_j^{(w)2}(\gamma) \hat  v_j(\gamma) x_j x_j^\T$ with
\begin{align*}
\hat  v_j(\gamma) & = \frac{ \hat p_{11j}\hat p_{12j}}{N_{1j}-1}  + \phi^2_j(\gamma)  \frac{ \hat p_{21j} \hat p_{22j}}{N_{2j}-1} , \quad j=1,\ldots, J.
\end{align*}
Our model-based variance estimator is
$\hat  V^{(wb)} = \hat D^{(w)-1}(\gamma) \hat C^{(wb)}(\gamma) \hat D^{(w)-1}(\gamma) |_{\gamma= \hat \gamma^{(w)}}$,
where $\hat C^{(wb)} (\gamma) = N_\bullet^{-1} \sum_{j=1}^J  q_j^{(w)2}(\gamma) \hat  v_j^{(b)}(\gamma) x_j x_j^\T$ with
\begin{align*}
\hat  v_j^{(b)} (\gamma) & =  \phi_j(\gamma) \left( \frac{ \hat p_{12j} \hat p_{21j} }{N_{1j}}  +   \frac{ \hat p_{11j}  \hat p_{22j}}{N_{2j}} \right) , \quad j=1,\ldots, J.
\end{align*}
The following properties can be established.

\begin{pro} \label{pro:pr-var-est}
Let $ v_j(\gamma) = \var \{ \hat p_{11j} - \phi_j (\gamma) \hat p_{21j} \}$.\\
(i) $ \hat  v_j^{(b)}(\gamma^*)$ is an unbiased estimator of $ v_j(\gamma^*)$ for $j=1,\ldots,J$, and $\hat  V^{(wb)}$ is a consistent estimator of $ V^{(w)}$ in both Settings I and II,
provided that model (\ref{eq:pr-model}) is valid. \\
(ii) Assume that $N_{1j} \ge 2$ and $N_{2j} \ge 2$ for each $j=1,\ldots,J$.  Then $ \hat  v_j(\gamma)$ is an unbiased estimator of $ v_j(\gamma)$ for any fixed $\gamma$ and $j=1,\ldots,J$.
Moreover, $\hat  V^{(w)}$ is a consistent estimator of $ V^{(w)}$ in both Settings I and II, with possible misspecification of model (\ref{eq:pr-model}).
\end{pro}

The estimator $\hat  v_j (\gamma)$ is based on the usual variance estimator for sample proportions with binary data.
Algebraically, $\hat  v_j (\gamma)$ and $ \hat  v_j^{(b)} (\gamma)$ are bivariate polynomials of $(\hat p_{11j}, \hat p_{21j})$, but $ \hat  v_j^{(b)} (\gamma)$ involves
only the cross-product $\hat p_{11j} \hat p_{21j}$, not $(\hat p_{11j}^2, \hat p_{21j}^2)$.
There is, in general, no definite direction in which of $\hat  v_j (\gamma)$ and $ \hat  v_j^{(b)} (\gamma)$ is greater, and hence
comparison of magnitudes between $\hat  V^{(w)}$ and $\hat  V^{(wb)}$ is problem-dependent.

\begin{rem} \label{rem:breslow-var}
The model-based variance estimator $N_\bullet^{-1} \hat  V^{(wb)}$ is, in the order on variance matrices, always no greater than $N_\bullet^{-1}\hat D^{(w)-1}(\hat\gamma^{(w)}) $,
which is the inverse negative Hessian of the criterion function (\ref{eq:pr-WEE-loss}), commonly reported as the
model-based variance estimator for the Breslow--Peto estimator $\hat\gamma^{(w)}$.
This can be shown by noting that $\hat D^{(w)} (\gamma)$ is identical to $\hat C^{(wb)} (\gamma)$ with $\hat  v_j^{(b)} (\gamma) $ replaced by
$\phi_j(\gamma) ( \hat p_{21j} / N_{1j} + \hat p_{11j}/N_{2j})$, which is at least as large as  $\hat  v_j^{(b)} (\gamma) $.
On the other hand, in the special case where $n_{11j}=0$ or $n_{21j}=0$ for $j=1,\ldots,J$ (for example, a total of one success in each table),
$\hat D^{(w)} (\gamma)$ is identical to $ \hat C^{(wb)} (\gamma)$ and hence the variance estimate $N_\bullet^{-1} \hat  V^{(wb)}$ numerically agrees with the commonly reported
variance estimate $N_\bullet^{-1} \hat D^{(w)-1}(\hat\gamma^{(w)}) $.
\end{rem}

\begin{rem} \label{rem:oddr-var-special}
Consider the special case where $n_{11j}=0$ or $n_{21j}=0$ for $j=1,\ldots,J$ (including but not restricted to a total of one success in each table).
The two estimates $\hat\beta^{(w)}$ and $\hat\gamma^{(w)}$ are identical to each other, as seen from the second equality in (\ref{eq:one-success}).
The matrix $\hat H^{(w)}(\beta)$ can also to shown to be identical to $\hat D^{(w)} (\gamma)$ and $\hat C^{(wb)} (\gamma)$,
but $\hat G^{(wb)}(\beta)$ does not seem to be related to $\hat C^{(wb)} (\gamma)$ in a simple way, with both $\beta$ and $\gamma$ evaluated at $\hat\beta^{(w)} =\hat\gamma^{(w)}$.
Hence the model-based variance estimate $N_\bullet^{-1} \hat  \Sigma^{(wb)}$ for $\hat\beta^{(w)}$ may differ from
$N_\bullet^{-1} \hat  V^{(wb)}$ and $N_\bullet^{-1} \hat D^{(w)-1}(\hat\gamma^{(w)}) $ for $\hat\gamma^{(w)}$, even though $\hat\beta^{(w)} = \hat\gamma^{(w)}$.
\end{rem}

\begin{rem} \label{rem:oddr-var-special}
We mainly study unconditional inference, with possible model misspecification. 
Conditional inference has been well established given four marginals in each table under odds ratio model (\ref{eq:oddr-model}) (Breslow 1981; McCullagh \& Nelder 1989),
although how to perform conditional inference under probability ratio model (\ref{eq:pr-model}) is not clear.
In our approach, both $\hat\beta^{(w)} $ and $\hat\gamma^{(w)}$ are constructed to
coincide with the conditional likelihood estimator $\hat\beta^{(c)}$ given a total of one success in each table under model (\ref{eq:oddr-model}).
Moreover, if model (\ref{eq:oddr-model}) is valid, then given a total of one success in each table, $N_\bullet^{-1} \hat  \Sigma^{(wb)}$ is a consistent variance estimator for
$\hat\beta^{(w)} = \hat\gamma^{(w)}$ by Robins et al.~(1986, Section 5), and so is the standard variance estimator $N_\bullet^{-1} \hat D^{(w)-1}(\hat\gamma^{(w)}) $ or $N_\bullet^{-1} \hat  V^{(wb)}$.
An interesting difference, however, is that $N_\bullet^{-1} \hat  V^{(wb)}$ and $N_\bullet^{-1} \hat D^{(w)-1}(\hat\gamma^{(w)}) $, but not $N_\bullet^{-1} \hat  \Sigma^{(wb)}$,
depend on the binary data $\{n_{11j}:j=1,\ldots,J\}$ only through the point estimate  $\hat\beta^{(w)} = \hat\gamma^{(w)}$.
\end{rem}

\begin{rem} \label{rem:log-rank}
Based on point and variance estimation, Wald and score-type tests can be derived by standard arguments (e.g., Lin \& Wei 1989).
Of particular interest is to test the null hypothesis that $p_{11j} = p_{21j}$ for $j=1,\ldots,J$, which can be expressed as $\beta^*=0$ in model (\ref{eq:oddr-model})
or equivalently $\gamma^*=0$ in model (\ref{eq:pr-model}), with constant covariates. A score-type test statistic based on $\zeta^{(w)}(\gamma)$ is
$\zeta^{(w)} (0) / \{ \sum_{j=1}^J q_j^{(w)2}(0) \hat v_j^{(b)}(0) \}^{1/2}$.
Under the null hypothesis, the variance estimator $\hat v_j^{(b)}(0)$ can be replaced by another unbiased estimator of $\var(\hat p_{11j} - \hat p_{21j})$ defined as
$$
\hat v^{(po)}_j = \hat p_{\cdot 1j} \hat p_{\cdot 2j} \frac{N_{1j}+N_{2j}}{N_{1j}+N_{2j}-1} \left(\frac{1}{N_{1j}} + \frac{1}{N_{2j}} \right),
$$
where $\hat p_{\cdot 1j}  = (p_{11j} + p_{21j})/(N_{1j} + N_{2j})$ and $\hat p_{\cdot 2j}=1-\hat p_{\cdot 1j}$ are the pooled-sample probability estimates.
The resulting test statistic can be directly shown to be
\begin{align*}
\frac{\zeta^{(w)} (0) } {\{ \sum_{j=1}^J q_j^{(w)2}(0) \hat v_j^{(po)}(0) \}^{1/2}}
= \frac{\sum_{j=1}^J (n_{11j} N_{2j} - n_{21j} N_{1j})/(N_{1j}+N_{2j}) }{\{\sum_{j=1}^J N_{1j} N_{2j}  \hat p_{\cdot 1j} \hat p_{\cdot 2j} /(N_{1j}+N_{2j}-1) \}^{1/2} },
\end{align*}
which is identical to the log-rank test statistic.
Similarly, the score-type test statistic
$\tau^{(w)} (0) / \{ \sum_{j=1}^J q_j^{(w)2}(0) \hat \sigma_j^{(b)}(0) \}^{1/2}$
also leads to the log-rank test statistic if $\hat \sigma_j^{(b)}(0)$ is replaced by the pooled-sample variance estimator $\hat v_j^{(po)}(0)$.
These connections are reassuring, especially for two-sample survival analysis in Section~\ref{sec:survival}, where the log-rank test is known to be applicable to survival data
in both continuous and discrete time.
\end{rem}

\section{Two-sample survival analysis} \label{sec:survival}

Suppose that survival data and covariates are obtained as $\{(Y_i, \delta_i, Z_i): i=1,\ldots,N\}$ from $N$ individuals,
where $Y_i = \min(T_i,C_i)$, $\delta_i = 1\{T_i \le C_i\}$, $T_i$ is an event time such as death time, $C_i$ is a censoring time, and $Z_i$ is a covariate.
All individuals are assumed to be event-free at entry, $T_i>0$ for all $i$, and hence it is possible that $Y_i=0$ if $\delta_i=0$, but not if $\delta_i=1$.
For two-sample analysis, each covariate $Z_i$ is two-level, being 1 or 2 if $i$th individual is from group 1 or 2.
Assume that $\{(T_i,C_i,Z_i): i=1,\ldots,N\}$ are independent and identically distributed copies of $(T,C,Z)$, and hence
$\{(Y_i,\delta_i,Z_i): i=1,\ldots,N\}$ are independent and identically distributed copies of $(Y,\delta,Z)$ with $Y = \min(T,C)$ and $\delta = 1\{T \le C\}$.
In addition, assume that the censoring and event variables, $C$ and $T$, are independent conditionally on the covariate $Z$.

In practice, time is usually recorded in discrete units such as days or weeks.
Assume that there is a discrete time scale, $0 = t_0 < t_1 < \cdots < t_J < t_{J+1}$, such that $(Y,\delta)$ and $(T,C)$ are properly
defined with $C \in \{t_0, t_1, \ldots, t_J\}$ and $T \in \{t_1, \ldots, t_J, t_{J+1}\}$ and the conditionally independent censoring assumption is satisfied.
There are subtle issues when the  survival data are collected by grouping continuous or fine-scaled measurements, but such detailed data are not available.
See Kaplan \& Meier (1958) and Thompson (1977) for early treatment and the Supplement for further discussion.

For survival analysis, it is commonly of interest to estimate hazard functions. In discrete time, the hazard at time $t_j$ given covariate $Z=z$ is defined as
the probability $ \pi_{z1j}  =  P( T =t_j | T \ge t_j , Z=z)$ for $z=1,2$.
Under conditionally independent censoring, $\pi_{z1j}$ can be identified from observed data as $\pi_{z1j}=p_{z1j}$, where $p_{j z}$ is the event probability at $t_j$ calculated within
the population risk set $\{ Y \ge t_j \}$,
\begin{align}
p_{z1j}  =  P( Y=t_j, \delta=1 | Y \ge t_j,Z=z ) . \label{eq:hazard}
\end{align}
The population risk set $\{Y \ge t_j \}$ represents individuals who are event-free (or alive) just prior to time $t_j$.
For ease of interpretation, we treat $(p_{11j}, p_{21j})$ interchangeably with the hazard probabilities  $(\pi_{11j}, \pi_{21j})$ whenever possible.
We also study inference about odds and probability ratios directly in terms of $(p_{11j}, p_{21j})$,
which coincide with $(\pi_{11j}, \pi_{21j})$ if  conditionally independent censoring holds, but otherwise remain empirically identifiable.
Our results are then applicable even if conditionally independent censoring is violated; see Tan (2006, Section 4) for a similar approach.

With the concept of risk sets, two-sample survival analysis can be easily related to analysis of $2 \times 2$ tables in Section~\ref{sec:2by2}.
For $z=1,2$ and $j=1,\ldots,J$, let $N_{zj}$ be the size of $\{1\le i \le N: Y_i \ge t_j, Z_i = z\}$, the sample risk set associated with time $t_j$ given covariate $Z=z$,
and let $n_{z1j}$ be the number of events (or deaths) at $t_j$ within the risk set, i.e., the size of $\{1\le i \le N: Y_i =t_j, \delta_i=1, Z_i = z\}$.
Then the survival data $\{(Y_i,\delta_i): i=1,\ldots,N\}$ can be transformed into a series of $J$ tables as shown in Table~\ref{tab:2by2}.
By this connection, it is helpful to exploit similar methods from analysis of $2\times 2$ tables for two-sample survival analysis.
Unless otherwise stated, we use the same notation $(p_{11j},p_{21j})$ and $(n_{11j},n_{21j})$ etc, although there are important differences.
In particular, the $J$ tables are not independent and the sizes $(N_{1j}, N_{2j})$ are random.

Consider two types of models for the probabilities $(p_{11j}, p_{21j})$.
The first is  model (\ref{eq:oddr-model}), $\log(\psi^*) = x_j^T \beta^*$, on
the odds ratios $\psi^*_j = p_{11j} p_{22j} / (p_{12j} p_{21j})$, that is,
\begin{align}
\frac{p_{11j}}{1-p_{11j}} = \exp (x_j^\T \beta^*) \frac{p_{21j}}{1-p_{21j}}, \quad j=1,\ldots,J. \label{eq:oddr-model2}
\end{align}
where $x_j = x(t_j)$ and each component of $x(\cdot)$ is a function of time, for example, a piecewise-constant function.
The second is model (\ref{eq:pr-model}), $\log(\phi^*) = x_j^T \gamma^*$, on
the probability ratios $\phi^*_j = p_{11j}/ p_{21j}$, that is,
\begin{align}
 p_{11j}  = \exp (x_j^\T \gamma^*) p_{21j} , \quad j=1,\ldots,J.\label{eq:pr-model2}
\end{align}
The first model (\ref{eq:oddr-model2}) is known as the discrete-time version of Cox's (1972) proportional hazards model, typically used when handling tied death times.
For this model, partial likelihood inference can be performed given the total numbers of deaths in the $J$ tables, but is computationally costly.
In practice, the Breslow--Peto approximation is widely employed, although it can be less accurate than other options (Efron 1977).
By comparison, the second model (\ref{eq:pr-model2}) has received limited attention, but seems more suitable to be called a proportional hazards model because it is directly concerned with the hazard ratios
$p_{11j}/ p_{21j}$.
Nevertheless, a remarkable finding which motivates our interest in model (\ref{eq:pr-model2}) is that the Breslow--Peto approximation to the exact partial likelihood estimator of $\beta^*$ in model (\ref{eq:oddr-model2})
yields a consistent estimator of $\gamma^*$ in model (\ref{eq:pr-model2}).

\subsection{Point estimation}

For models (\ref{eq:oddr-model2}) and (\ref{eq:pr-model2}), point estimators of $\beta^*$ and $\gamma^*$ can be directly adopted from Section~\ref{sec:2by2}
by the transformation of the survival data into $2\times 2$ tables.
Such connections are in fact exploited in the construction of estimators in Section~\ref{sec:2by2}.

For model (\ref{eq:oddr-model2}), the weighted Mantel--Haenszel estimator $\hat\beta^{(w)}$ is a solution to $\tau^{(w)} (\beta)=0$
with $\tau^{(w)} (\beta)$ defined in (\ref{eq:WEE}). Computation of $\hat\beta^{(w)}$ is straightforward due to the simplicity of the estimating function $\tau^{(w)}(\beta)$.
By comparison, the exact partial likelihood estimator $\hat \beta^{(c)}$ is a solution to $s(\beta)=0$,
with $s(\beta)$ defined in (\ref{eq:cond-score}).
Computation of $\hat\beta^{(c)}$ is  burdensome because the conditional mean $\mu_{11j}(\beta)$ is intractable when the total number of deaths, $n_{11j}+n_{21j}$, is large.
The widely used Breslow--Peto approximation to the partial likelihood leads to the estimating function
\begin{align*}
\sum_{j=1}^J \sum_{1\le i \le N : Y_i=t_j, \delta_i= 1}  \left\{ 1\{Z_i=1\} - \frac{N_{1j} \psi_j(\beta) }{N_{1j} \psi_j(\beta) + N_{2j} } \right\} x_j .
\end{align*}
By the definitions of $(n_{11j},n_{21j})$, this function can be calculated as
\begin{align*}
\sum_{j=1}^J \left\{ n_{11j}- \frac{(n_{11j}+n_{21j})N_{1j} \psi_j(\beta) }{ N_{1j} \psi_j(\beta) + N_{2j} } \right\} x_j ,
\end{align*}
which coincides with $\zeta^{(w)}(\gamma)$ in (\ref{eq:pr-WEE}) and (\ref{eq:pr-WEE2}),
except for the change between $\beta$ and $\gamma$, with $\psi_j(\beta) = \exp(x_j^\T\beta)$ and $\phi_j(\gamma) = \exp(x_j^\T \gamma)$ as defined in Section~\ref{sec:2by2}.

\begin{pro} \label{pro:breslow-peto}
The estimator defined by the Breslow--Peto approximation to the partial likelihood estimator for $\beta^*$ in odds ratio model (\ref{eq:oddr-model2})
is algebraically identical to the estimator $\hat \gamma^{(w)}$ for $\gamma^*$ in probability ratio model (\ref{eq:pr-model2}).
\end{pro}

The probability ratios $p_{11j}/p_{21j}$ are always closer to 1 than the odds ratios $p_{11j} p_{22j}/ $ $(p_{12j}p_{21j})$, except when $p_{11j}=p_{21j}$.
This result agrees with the previous finding that the Breslow--Peto approximation produces a conservative bias in estimating
regression coefficients too close to 0 in proportional hazards models (Cox \& Oaks 1984).

\subsection{Variance estimation}

We derive model-based and model-robust estimators of the asymptotic variance of the weighted  Mantel--Haenszel estimator $\hat \beta^{(w)}$ associated with
model (\ref{eq:oddr-model2}) or respectively the Breslow--Pero estimator $\hat \gamma^{(w)}$ associated with model (\ref{eq:pr-model2}).
As mentioned earlier, a complication is that the stratum sizes $(N_{1j}, N_{2j})$ are random and
the $2\times 2$ tables constructed from risk sets over time are correlated over time. However, this difficulty can be handled by deriving
appropriate influence functions, which take into account the contributions over time from each individual in the sample.

First, we describe the asymptotic distribution of $\hat \gamma^{(w)}$ and a model-robust variance estimator, while allowing for possible misspecification of model (\ref{eq:pr-model2}).
The following notation is used. For $z=1,2$, let
$ P_{zj} = P( Y \ge t_j, Z=z) $ and
$ P_{z1j} = P( Y= t_j, \delta=1, Z=z) $, which are the unconditional probabilities of
an individual being included in the risk set associated with time $t_j$ or respectively observed to experience an event at time $t_j$.
The hazard is then identified by $p_{z1j} = P_{z1j} / P_{zj}$ in (\ref{eq:hazard}) under conditionally independent censoring.
Denote the corresponding indicators as $ I_{zj} (Y,Z) = 1\{Y \ge t_j ,Z=z \}$
and
$ I_{z1j} (Y,\delta,Z) =  1\{Y = t_j , \delta = 1, Z=z \}$. Define
\begin{align*}
 h_j(Y,\delta,Z; \gamma) &= \frac{P_{2j}}{ P_{1j} \phi_j(\gamma) + P_{2j}} ( I_{11j} - p_{11j} I_{1j} ) - \frac{P_{1j}\phi_j(\gamma)}{ P_{1j} \phi_j(\gamma) + P_{2j}} ( I_{21j} - p_{21j} I_{2j} ) \\
& \quad +\frac{p_{11j} - \phi_j(\gamma) p_{21j}}{ \{P_{1j} \phi_j(\gamma) + P_{2j} \}^2} \{P_{2j}^2 I_{1j} + \phi_j(\gamma) P_{1j}^2 I_{2j} \} .
\end{align*}
For a column vector $b$, denote $b^{\otimes 2} =b b^\T$.
The matrix $N \hat D^{(w2)}$ is identical to $N_\bullet \hat D^{(w)}$ in Section~\ref{sec:pr-var},
whereas $\hat C^{(w2)}$ and $\hat V^{(w2)}$ differ from $\hat C^{(w)}$ and $\hat V^{(w)}$ even after rescaling.

\begin{pro} \label{pro:pr-asym}
Assume that $P(T > t_J) \ge p_0$ for a constant $p_0>0$. \\
(i) $\hat\gamma^{(w)}$ converges in probability to a target value $\bar\gamma^{(w2)}$ which solves the equation
\begin{align}
0 = \sum_{j=1}^J \frac{P_{1j} P_{2j}}{ P_{1j} \phi_j(\gamma) + P_{2j}}  \{p_{11j} - \phi_j(\gamma) p_{21j} \} x_j. \label{eq:pr-target}
\end{align}
Moreover, $N^{1/2} ( \hat\gamma^{(w)} - \bar\gamma^{(w2)})$ converges in distribution to $\N(0, V^{(w2)})$, with
$ V^{(w2)}= D^{(w2)-1}(\gamma)  C^{(w2)}(\gamma) D^{(w2)-1}(\gamma) |_{\gamma= \bar\gamma^{(w2)}}$, where
\begin{align*}
C^{(w2)} (\gamma) 
& = \var \left\{ \sum_{j=1}^J h_j(Y,\delta,Z; \gamma) x_j \right\},\\
D^{(w2)}  (\gamma) & = \sum_{j=1}^J  (P_{1j} p_{11j} + P_{2j} p_{21j} )\frac{P_{1j}P_{2j}  \phi_j(\gamma)}{ \{P_{1j} \phi_j(\gamma) + P_{2j}\}^2} x_j x_j^\T .
\end{align*}
(ii) A consistent estimator of $ V^{(w2)}$ is
$\hat V^{(w2)} = \hat D^{(w2)-1}(\gamma) \hat C^{(w2)}(\gamma) \hat D^{(w2)-1}(\gamma) |_{\gamma= \bar\gamma^{(w2)}}$,
where $ \hat C^{(w2)}(\gamma)  = N^{-1} \sum_{i=1}^N \{ \sum_{j=1}^J \hat h_j (Y_i, \delta_i, Z_i;\gamma) x_j \}^{\otimes 2}$,
$\hat h_j (Y, \delta, Z;\gamma)$ is defined as $h_j (Y, \delta,$ $ Z; \gamma)$, and $\hat D^{(w2) }(\gamma) $ defined as $ D^{(w2) }(\gamma) $,
with $(P_{1j}, P_{2j})$ replaced by $(\hat P_{1j},\hat P_{2j}) = (N_{1j},N_{2j})/N$ and $(p_{11j}, p_{21j})$ replaced by $(\hat p_{11j}, \hat p_{21j})$.
\end{pro}

As shown in the proof, $h_j(Y,\delta,Z; \gamma)$ is obtained by linearizing the function
$ \{ n_{11j} N_{2j} - \phi_j(\gamma) n_{21j} N_{1j}\} / \{N_{1j} \phi_j(\gamma) + N_{2j}\}$ as a sample average
$ \sum_{i=1}^N h_j(Y_i,\delta_i,Z_i; \gamma)$, with $\gamma$ evaluated at $\bar\gamma^{(w2)}$, which captures
the contribution of all individuals to the $j$th term of $\zeta^{(w)}(\gamma)$ in (\ref{eq:pr-WEE}), calculated from the risk set associated with time $t_j$.
If model (\ref{eq:pr-model2}) is correctly specified, then  $\bar\gamma^{(w2)}=\gamma^*$ and hence
$\hat \gamma^{(w)}$ is a consistent estimator of $\gamma^*$.
Moreover, the third term of $h_j(Y,\delta,Z; \gamma)$ reduces to 0 with $\gamma=\gamma^*$, because $p_{11j} - \phi_j(\gamma^*) p_{21j}=0$ in this case.
This simplification can also be seen from the fact that $ n_{11j} N_{2j} - \phi_j(\gamma^*) n_{21j} N_{1j}$, $j=1,\ldots,J$, forms a
martingale difference when successively conditioning on the $j$th risk set, and hence
the effect of the variation in the stratum sizes $(N_{1j}, N_{2j})$ is negligible, of the order $o_p(N^{-1/2})$.
If model (\ref{eq:pr-model2}) is misspecified, then $p_{11j} - \phi_j(\gamma) p_{21j} $ is in general nonzero with $\gamma=\bar\gamma^{(w2)}$ and
hence the third term of $h_j(Y,\delta,Z; \gamma)$ is needed.

The preceding discussion shows that if model (\ref{eq:pr-model2}) is correctly specified, then a consistent
estimator of $V^{(w2)}$ is $\hat V^{(w2a)}$, defined as $\hat V^{(w2)}$ except with the third term of $\hat h_j (Y, \delta, Z;\gamma)$ removed.
The variance estimator $N^{-1}\hat V^{(w2a)}$ is then identical to $N_\bullet^{-1}\hat V^{(w)}$ in Section~\ref{sec:pr-var} except for $(N_{1j}-1, N_{2j}-1)$ used in $\hat v_j(\gamma)$.
Such a small-sample adjustment is technically not needed here because $P(T > t_J)$ is assumed to be bounded away from 0. This boundedness condition
is standard in large sample theory for survival analysis (e.g., Anderson et al.~1993, Section IV.3.2), although further investigation can be of interest.
Alternatively, the preceding discussion also shows that another valid model-based variance estimator for $\hat\gamma^{(w)}$ is $N_\bullet^{-1} \hat V^{(wb)}$,
which, by Remark~\ref{rem:breslow-var}, can be directly compared with the existing model-based variance estimator for $\hat\gamma^{(w)}$.

\begin{cor} \label{cor:pr-correct-model}
(i) If model (\ref{eq:pr-model2}) is correctly specified, then $\hat \gamma^{(w)}$ is a consistent estimator of $\gamma^*$ with asymptotic variance $N^{-1} V^{(w2)}$,
where the third term of $h_j (Y, \delta, Z;\gamma)$ can be removed.
Moreover, $\hat V^{(w2b)}=(N/N_\bullet)\hat V^{(wb)}$, with $\hat V^{(wb)}$ defined in Section~\ref{sec:pr-var}, is a consistent estimator of $V^{(w2)}$.\\
(ii) The variance estimator $N^{-1}\hat V^{(w2b)}$ is, in the order on variances matrices, no greater than $N^{-1}\hat D^{(w2) -1} (\hat\gamma^{(w)})$, which is the commonly reported model-based variance estimator for
the Breslow--Peto estimator $\hat\gamma^{(w)}$ when used as an approximation to the exact partial likelihood estimator in model (\ref{eq:oddr-model2}).
\end{cor}

\begin{rem} \label{rem:lin-wei}
Proposition~\ref{pro:pr-asym} can be deduced from Lin \& Wei (1989), who derived a model-robust variance estimator for the partial likelihood estimator in Cox's
proportional hazards model. In the presence of tied death times, Lin \& Wei's result remains valid if the Breslow--Peto modification is used.
Their variance estimator for $\hat\gamma^{(w)}$ in the two-sample discrete-time setting can be expressed as $N^{-1} \hat V^{(w2)}$, with $\hat V^{(w2)}$ defined in Proposition~\ref{pro:pr-asym} but
$\hat C^{(w2)}(\gamma) =  N^{-1} \sum_{i=1}^N \{ \sum_{j=1}^J \hat h_j^{\mbox{\tiny LW}} (Y_i, \delta_i, Z_i;\gamma) x_j \}^{\otimes 2}$, where
\begin{align*}
\hat h_j^{\mbox{\tiny LW}}(Y,\delta,Z; \gamma) & = 1\{Y=t_j\} \delta \left\{1\{Z=1\} - \frac{ N_{1j} \phi_j(\gamma)}{N_{1j} \phi_j(\gamma) + N_{2j}} \right\} \\
& \quad - 1\{ Y \ge t_j\}  \frac{ (n_{11j}+n_{21j}) \phi_j(\gamma)^{1\{Z=1\}}}{N_{1j} \phi_j(\gamma) + N_{2j}}  \left\{1\{Z=1\}  - \frac{ N_{1j} \phi_j(\gamma)}{N_{1j} \phi_j(\gamma) + N_{2j}} \right\}.
\end{align*}
We show that $\hat h_j(Y,\delta,Z; \gamma)$ algebraically coincides with $\hat h_j^{\mbox{\tiny LW}}(Y,\delta,Z; \gamma)$ in the proof of Proposition~\ref{pro:pr-asym},
and hence our model-robust variance estimator for $\hat\gamma^{(w)}$ agrees with Lin \& Wei's (1989).
On one hand, this agreement is expected because there is a unique model-robust influence function for $\hat \gamma^{(w)}$, whether model (\ref{eq:oddr-model2}) or
(\ref{eq:pr-model2}) is valid or not. On the other hand, as discussed above, our representation with $\hat h_j(Y,\delta,Z; \gamma)$ is explicit in separating what variation persists
and what vanishes if model (\ref{eq:pr-model2}) is valid. Accounting for the former component leads to a new model-based variance estimator for $\hat\gamma^{(w)}$,
which is no greater than the usually used model-based variance estimator for $\hat\gamma^{(w)}$.
\end{rem}

Next, we describe the asymptotic distribution of $\hat \beta^{(w)}$ and a model-robust variance estimator, while allowing for possible misspecification of model (\ref{eq:oddr-model2}).
The same notation is used as above.  Define
\begin{align*}
g_j(Y,\delta,Z; \beta) &= \frac{P_{2j} \{ p_{22j} +  \psi_j(\beta) p_{21j}\}}{ P_{1j} \psi_j(\beta) + P_{2j}} ( I_{11j} - p_{11j} I_{1j} )
- \frac{P_{1j}\{ p_{11j} +  \psi_j(\beta) p_{12j}\} }{ P_{1j} \psi_j(\beta) + P_{2j}} ( I_{21j} - p_{21j} I_{2j} ) \\
& \quad +\frac{p_{11j}p_{22j} - \psi_j(\beta) p_{12j} p_{21j}}{ \{P_{1j} \psi_j(\beta) + P_{2j} \}^2} \{P_{2j}^2 I_{1j} + \psi_j(\beta) P_{1j}^2 I_{2j} \} .
\end{align*}
The matrix $N \hat H^{(w2)}$ is identical to $N_\bullet \hat H^{(w)}$ in Section~\ref{sec:oddr-var},
whereas $\hat G^{(w2)}$ and $\hat \Sigma^{(w2)}$ differ from $\hat G^{(w)}$ and $\hat \Sigma^{(w)}$ even after rescaling.

\begin{pro} \label{pro:oddr-asym}
Assume that $P(T > t_J) \ge p_0$ for a constant $p_0>0$. \\
(i) $\hat\beta^{(w)}$ converges in probability to a target value $\bar\beta^{(w2)}$ which solves the equation
\begin{align}
0 = \sum_{j=1}^J \frac{P_{1j} P_{2j}}{ P_{1j} \psi_j(\beta) + P_{2j}}  \{p_{11j}p_{22j} - \psi_j(\beta) p_{12j} p_{21j} \} x_j. \label{eq:oddr-target}
\end{align}
Moreover, $N^{1/2} ( \hat\beta^{(w)} - \bar\beta^{(w2)})$ converges in distribution to $\N(0, \Sigma^{(w2)})$, with
$ \Sigma^{(w2)}= H^{(w2)-1}(\beta) G^{(w2)}(\beta) H^{(w2)-1}(\beta) |_{\beta= \bar\beta^{(w2)}}$, where
\begin{align*}
G^{(w2)} (\beta)
& = \var \left\{ \sum_{j=1}^J g_j(Y,\delta,Z; \beta) x_j \right\},\\
H^{(w2)}  (\beta) & = \sum_{j=1}^J  (P_{1j} p_{11j}p_{22j} + P_{2j} p_{12j} p_{21j} )\frac{P_{1j}P_{2j}  \psi_j(\beta)}{ \{P_{1j} \psi_j(\beta) + P_{2j}\}^2} x_j x_j^\T .
\end{align*}
(ii) A consistent estimator of $ \Sigma^{(w2)}$ is
$\hat \Sigma^{(w2)} = \hat H^{(w2)-1}(\beta) \hat G^{(w2)}(\beta) \hat H^{(w2)-1}(\beta) |_{\beta= \bar\beta^{(w2)}}$,
where $ \hat G^{(w2)}(\beta)  = N^{-1} \sum_{i=1}^N \{ \sum_{j=1}^J \hat g_j (Y_i, \delta_i, Z_i;\beta) x_j \}^{\otimes 2}$,
$\hat g_j (Y, \delta, Z;\beta)$ is defined as $g_j (Y, \delta,$ $ Z; \beta)$, and $\hat H^{(w2) }(\beta) $ defined as $ H^{(w2) }(\beta) $,
with $(P_{1j}, P_{2j})$ replaced by $(\hat P_{1j},\hat P_{2j}) = (N_{1j},N_{2j})/N$ and $(p_{11j}, p_{21j})$ replaced by $(\hat p_{11j}, \hat p_{21j})$.
\end{pro}

The preceding results exhibit a remarkable similarity to Proposition~\ref{pro:pr-asym}.
If model (\ref{eq:oddr-model2}) is correctly specified, then  $\bar\beta^{(w2)}=\beta^*$ and hence
$\hat \beta^{(w)}$ is a consistent estimator of $\beta^*$.
Moreover, the third term of $g_j(Y,\delta,Z; \beta)$ reduces to 0 with $\beta=\beta^*$.
Hence $\hat\Sigma^{(w2)}$ can be simplified as $\hat \Sigma^{(w2a)}$, defined as
$\hat \Sigma^{(w2)}$ with the third term of $g_j(Y,\delta,Z; \beta)$ removed.
The variance estimator $N^{-1}\hat \Sigma^{(w2a)}$ for $\hat\beta^{(w)}$ is then identical to $N_\bullet^{-1}\hat \Sigma^{(w)}$ in Section~\ref{sec:oddr-var}
except with $\hat \sigma_j(\gamma)$ modified such that its third term is removed and $(N_{1j},N_{2j})$ are used instead of $(N_{1j}-1, N_{2j}-1)$.
Alternatively, another valid model-based variance estimator for $\hat\beta^{(w)}$ is
$N_\bullet^{-1} \hat\Sigma^{(wb)}$ derived from Robins et al.~(1986) in Remark~\ref{rem:robins-var}.

\begin{cor} \label{cor:oddr-correct-model}
If model (\ref{eq:oddr-model2}) is correctly specified, then $\hat \beta^{(w)}$ is a consistent estimator of $\beta^*$ with asymptotic variance $N^{-1} \Sigma^{(w2)}$,
where the third term of $g_j (Y, \delta, Z;\gamma)$ can be removed.
Moreover, $\hat \Sigma^{(w2b)}=(N/N_\bullet)\hat \Sigma^{(wb)}$, with $\hat \Sigma^{(wb)}$ defined in Section~\ref{sec:oddr-var}, is a consistent estimator of $\Sigma^{(w2)}$.
\end{cor}

So far, our development assumes that the survival data are recorded in pre-specified intervals, as commonly found in practice.
From a theoretical perspective, consider the setting where the survival time $T$ is absolutely continuous with hazard functions $\lambda(t| Z)$.
Both models (\ref{eq:oddr-model2}) and (\ref{eq:pr-model2}) can be seen to reduce to
a Cox proportional hazards model
\begin{align}
\lambda(t| Z=1) = \lambda (t| Z=0) \exp\{ \alpha^{*\T} x(t)\} , \label{eq:cox-model}
\end{align}
where $\alpha^*$ is an unknown coefficient vector.
A natural extension is then to apply the proposed point and variance estimators, with time intervals chosen to be sufficiently small such that there is a total of at most one death at $t_j$ in each interval $(t_{j-1},t_j]$.
By construction, both $\hat\beta^{(w)}$ and $\hat\gamma^{(w)}$ coincide with the conditional likelihood estimator $\hat \beta^{(c)}$, i.e.,
the standard partial likelihood estimator of $\alpha^*$ without tied death times.
Moreover, by Remark~\ref{rem:breslow-var}, the model-based variance estimator $N^{-1}\hat V^{(w2b)}$ for $\hat\gamma^{(w)}$
is identical to the usual model-based variance estimator $N^{-1} \hat D^{(w2)-1} (\hat\gamma^{(w)})$ for $\hat\beta^{(c)}$.
By Remark~\ref{rem:oddr-var-special} and Robins et al.~(1986, Section 5), the model-based variance estimator $N^{-1}\hat \Sigma^{(w2b)}$, even though different from $N^{-1} \hat D^{(w2)-1} (\hat\gamma^{(w)})$,
is also a consistent variance estimator for $\hat\beta^{(c)}$ if model (\ref{eq:cox-model}) is correctly specified.
Finally, by Remark~\ref{rem:lin-wei}, the model-robust variance estimator $N^{-1}\hat V^{(w2)}$ for $\hat\gamma^{(w)}$ coincides with
Lin \& Wei's (1989) variance estimator. Such a relationship does not hold for the model-robust variance estimator $N^{-1}\hat \Sigma^{(w2)}$ for $\hat\beta^{(w)}$.
From these remarks, we obtain the following result.

\begin{cor} \label{cor:cox-continuous}
If model (\ref{eq:cox-model}) is correctly specified, then $\hat\beta^{(w)}=\hat \gamma^{(w)}$ is a consistent estimator of $\alpha^*$ with an asymptotic variance matrix which can be
consistently estimated by $N^{-1}\hat \Sigma^{(w2b)}$ and $N^{-1}\hat V^{(w2b)}$. Moreover, the asymptotic variance matrix of $\hat \gamma^{(w)}$ can be consistently estimated by $N^{-1}\hat V^{(w2)}$,
with possible misspecification of model (\ref{eq:cox-model}).
\end{cor}

Although both models (\ref{eq:oddr-model2}) and (\ref{eq:pr-model2}) reduce to model (\ref{eq:cox-model}) in the continuous-time limit,
the estimator $\hat\gamma^{(w)}$ and its model-based and model-robust variance estimators $N^{-1}\hat V^{(w2b)}$ and $N^{-1}\hat V^{(w2)}$ associated with model (\ref{eq:pr-model2})
extend directly from discrete time to standard partial likelihood inference in the continuous-time limit.
For model (\ref{eq:oddr-model2}), model-robust variance estimation seems not directly extended for the estimator $\hat\beta^{(w)}$ in the continuous-time limit, and also remains to be studied for
the conditional likelihood estimator $\hat\beta^{(c)}$.
Further investigation is needed to fully address questions.

\section{Numerical studies}

\subsection{Analysis of $2 \times 2$ tables}

First, we reanalyze the data from the Oxford Childhood Cancer Survey, as described in Breslow \& Day (1980). The survey is
a retrospective study in which cases (children who died of cancer) and controls (children who were alive and well) were identified,
and their exposure to in utero radiation were ascertained. Hence the factor is dying of cancer or not, and the response is radiation exposure or not.
A total of 120 strata were used by age and year of birth. Following previous analyses,
we fit three odds ratio models: (i) $\log (\psi_j) = \beta_0$, (ii) $\log (\psi_j) = \beta_0 + \beta_1 x_j$, and (iii)  $\log (\psi_j) = \beta_0 + \beta_1 x_j + \beta_2 (x_j^2-22) $,
where $x_j$ indexes year of birth.
Table~\ref{tab:oxford} presents the point estimates and associated standard errors.
The weighted Mantel--Haenszel estimates and standard errors agree well with those from conditional likelihood inference,
more closely than Davis's (1985) results.
The (unweighted) Mantel--Haenszel estimates differ noticeably from conditional likelihood estimates, sometimes with inflated standard errors.
The model-based and model-robust standard errors appear to be aligned with each other here.

\begin{table}
\caption{Comparison of estimates from the Oxford Childhood Cancer Survey} \label{tab:oxford}
\small
\begin{center}
\begin{tabular*}{.9\textwidth}{ll rrr rrr rrr} \hline
     && $\hat\beta_0$  & bSE  & rSE & $\hat\beta_1$ & bSE  & rSE  & $\hat\beta_2$ & bSE  & rSE  \\ \hline
MH   && $.5051$ & $.0561$  & $.0561$  & \\
wMH  && $.5051$ & $.0561$  & $.0561$  & \\
CML  && $.5051$ & $.0561$  & NA \\ \cline{3-11}
MH   && $.5180$ & $.0565$  & $.0565$  & $-.0428$ & $.0162$ & $.0158$ \\
wMH  && $.5166$ & $.0564$  & $.0564$  & $-.0383$ & $.0143$ & $.0140$ \\
CML  && $.5165$ & $.0564$  & NA       & $-.0385$ & $.0144$   \\ \cline{3-11}
MH   && $.5694$ & $.0616$  & $.0614$  & $-.0462$ & $.0154$ & $.0153$  & $.0073$ & $.0032$ & $.0031$ \\
wMH  && $.5646$ & $.0607$  & $.0605$  & $-.0445$ & $.0149$ & $.0148$  & $.0067$ & $.0029$  & $.0029$ \\
CML  && $.5645$ & $.0608$  & NA       & $-.0445$ & $.0149$ &  NA      & $.0067$ & $ .0030$ & NA \\ \hline
\end{tabular*} \\[1ex]
\parbox{.9\textwidth}{\scriptsize Note: MH, wMH, and CML denote $\hat\beta^{(0)}$, $\hat\beta^{(w)}$, and $\hat\beta^{(c)}$ in Section~\ref{sec:2by2}.
bSE and rSE are model-based and model-robust standard errors. The results for CML are taken from Breslow \& Cologne (1986, Table 3).}
\end{center} \vspace{-.2in}
\end{table}

To further compare estimators, we conduct simulations under various settings similar to previous studies (e.g., Hauck et al.~1982; Robins et al.~1986). In particular,
Table~\ref{tab:2by2-sim} presents results from 2000 repeated simulations in the following two settings, where the success probabilities $(p_{11j}, p_{21j})$ are close to 0. See the Supplement for additional results.
For the first setting, $J=40$ tables are simulated with log odds ratios $\psi_j=\log(2)$, probabilities $p_{21j} = .03 + .001 j$ between $.031$ and $.07$
for $j=1,\ldots, 40$, and binomial sizes $(N_{1j}, N_{2j})=(16,4)$ for $j=1,\ldots,20$
and $(4,16)$ for $j=21, \ldots, 40$.
The second setting is the same as the first, except that $p_{11j}$ and $p_{21j}$ are related with log probability ratios $\phi_j=\log(2)$ for $j=1,\ldots,40$.
For each setting, models (\ref{eq:oddr-model}) and (\ref{eq:pr-model}) are fit with constant covariates, corresponding to common odds ratios or probability ratios,
which are valid only in the first or second setting respectively.

\begin{table}
\caption{Comparison of estimates from simulations of $2 \times 2$ tables} \label{tab:2by2-sim}
\small
\begin{center}
\begin{tabular*}{.9\textwidth}{ll rrrr ll rrrr} \hline
     && Point   & SD       & bSE      & rSE     &&        & Point  &  SD  &  bSE      & rSE \\ \hline
     && \multicolumn{10}{c}{log odds ratio $=\log(2) = .6931$} \\
MH   && $.7034$ & $.3581$  & $.3651$  & $.3550$ &&        &  \\
wMH  && $.6936$ & $.3465$  & $.3509$  & $.3509$ &&  BP    & $.6376$  & $.3157$ & $.3212$ & $.3195$ \\
CML  && $.6924$ & $.3453$  & $.3503$  & NA      && oldBP  & $.6376$  & $.3156$ & $.3349$ & NA \\  \cline{2-12}
     && \multicolumn{10}{c}{log probability ratio $=\log(2) = .6931$} \\
MH   && $.7616$ & $.3556$  & $.3611$  & $.3525$ &&        &  \\
wMH  && $.7536$ & $.3448$  & $.3475$  & $.3487$ &&  BP    & $.6907$  & $.3126$ & $.3169$ & $.3162$ \\
CML  && $.7519$ & $.3433$  & $.3468$  & NA      && oldBP  & $.6907$  & $.3126$ & $.3310$ & NA \\  \hline
\end{tabular*} \\[1ex]
\parbox{.9\textwidth}{\scriptsize Note: Point and SD are the Monte Carlo mean and standard deviation of the point estimates, and bSE and rSE
are the square roots of the Monte Carlo mean of the model-based and model-robust variance estimates. BP denotes $\hat \gamma^{(w)}$
with variance estimates $N_\bullet^{-1} \hat V^{(wb)}$ and $N_\bullet^{-1} \hat V^{(w)}$.
CMLE or oldBP denotes conditional logistic regression \texttt{clogit} with exact calculation or Breslow--Peto approximation in R package \texttt{survival}.}
\end{center} \vspace{-.2in}
\end{table}

The following observations can be obtained from Table~\ref{tab:2by2-sim}.
In the first setting, the weighted Mantel--Haenszel (wMH) and conditional maximum likelihood (CML) estimators perform similarly to each other, with smaller biases and variances than
unweighted Mantel--Haenszel (MH). The Breslow--Peto (BP) estimator, by our calculation or equivalently R package \texttt{survival}, is downward biased compared with the true log odds ratio $\log(2)$,
because BP can be seen to be centered around a target log probability ratio which is smaller than the common log odds ratio.
The model-based variance estimator commonly reported for BP is biased upward, whereas the proposed model-robust variance estimator
(as well as the new model-based variance estimator, although not guaranteed by theory) reasonably matches the Monte Carlo standard deviation.

In the second setting, BP is centered around the true log probability ratio with a negligible bias as expected. The commonly used variance estimator for BP remains biased upward,
where both the proposed model-based and model-robust variance estimators agree properly with the Monte Carlo standard deviation.
The estimators, MH, wMH, and CML, are upward biased compared with the true log probability ratio $\log(2)$,
because these estimators can be considered to be centered about a target log odds ratio which is greater than the common log probability ratio.

\subsection{Two-sample survival analysis}

First, we perform two-sample analysis of the data on a Veteran's Administration lung
cancer trial used in Kalbfleisch and Prentice (1980).
The data consist of 137 observations with right-censored survival time (apparently in days) and several covariates.
For two-sample analysis, the two groups are defined by a treatment variable $Z$, labeled as 1 or 2 if test or standard.
Kaplan--Meier survival curves suggest non-proportional hazards over time in the two groups (see the Supplement).
Hence we fit odds ratio model (\ref{eq:oddr-model2}) and probability ratio model (\ref{eq:pr-model2}),
with the time functions $x(t)=(1, x_1(t), x_2(t))^\T$, where $x_1(t)=1\{100 < t\le 200\}$ and $x_2(t)=1\{t>200\}$.
In addition, to study discrete-time inference, we also apply various estimators to further discretized data, obtained by grouping the original times in intervals of 20 days.
For concreteness, the censored-late option is used as discussed in the Supplement, i.e., an uncensored time in $(t_{j-1},t_j]$ is labeled $t_j$, whereas
a censored time in $[t_{j-1},t_j)$ is labeled $t_j$.

Table~\ref{tab:VA} presents the results on the original and discretized data.
For the original data with a small number of tied deaths, all the estimates obtained are similar to each other in various degrees, although
the BP point estimates are slightly closer to 0 than the estimates from MH, wMH, and CML,
and the model-based variance estimates on the row oldBP are larger than those on the row BP as expected by Corollary~\ref{cor:pr-correct-model}.
For the discretized data with more tied deaths, the BP point estimates
are more substantially closer to 0 than the estimates from MH, wMH, and CML, which remain relatively similar to each other.
This difference can be properly explained by the fact that BP estimates are associated with odds ratios,
whereas the other estimates are associated with probability ratios,
rather than poor approximation of BP to CML as would often be claimed.
The commonly reported model-based variance estimates on the row oldBP are also more inflated compared with the proposed variance estimates on the row BP.

\begin{table}
\caption{Comparison of estimates from veteran's lung cancer data} \label{tab:VA}
\small
\begin{center}
\begin{tabular*}{.95\textwidth}{ll rrr rrr rrr} \hline
     && $\hat\beta_0$  & bSE  & rSE & $\hat\beta_1$ & bSE  & rSE  & $\hat\beta_2$ & bSE  & rSE  \\ \hline
     && \multicolumn{9}{c}{Original data} \\
MH   && $.3989$ & $.2282$  & $.2268$  & $-1.1440$ & $.5019$ & $.4957$  & $-.9554$ & $.5376$ & $.5058$ \\
wMH  && $.3996$ & $.2286$  & $.2286$  & $-1.1399$ & $.4991$ & $.4972$  & $-.9433$ & $.5273$ & $.5019$ \\
CML  && $.3996$ & $.2288$  & NA       & $-1.1439$ & $.5002$ &  NA      & $-.9433$ & $.5287$ & NA \\
oldBP&& $.3960$ & $.2277$  & $.2265$  & $-1.1363$ & $.4988$ & $.4962$  & $-.9396$ & $.5283$ & $.5009$ \\
BP   && $.3960$ & $.2267$  & $.2265$  & $-1.1363$ & $.4984$ & $.4962$  & $-.9396$ & $.5278$ & $.5009$ \\  \cline{3-11}

     && \multicolumn{9}{c}{Discretized data} \\
MH   && $.4270$ & $.2494$  & $.2479$  & $-1.1969$ & $.5291$ & $.5312$  & $-1.0322$ & $.5634$ & $.5468$ \\
wMH  && $.4292$ & $.2507$  & $.2512$  & $-1.2020$ & $.5311$ & $.5372$  & $-1.0291$ & $.5598$ & $.5470$ \\
CML  && $.4258$ & $.2491$  & NA       & $-1.1902$ & $.5284$ &  NA      & $-1.0054$ & $.5549$ & NA \\
oldBP&& $.3541$ & $.2275$  & $.2080$  & $-1.0361$ & $.4985$ & $.4744$  & $-.8881$ & $.5279$ & $.4822$ \\
BP   && $.3541$ & $.2070$  & $.2080$  & $-1.0361$ & $.4684$ & $.4744$  & $-.8881$ & $.4953$ & $.4822$ \\  \hline
\end{tabular*} \\[1ex]
\parbox{.95\textwidth}{\scriptsize Note: MH and wMH denote $\hat\beta^{(0)}$ and $\hat\beta^{(w)}$ adopted from Section~\ref{sec:2by2},
but with variance estimation in Section~\ref{sec:survival}.
BP denotes $\hat \gamma^{(w)}$ with variance estimates $N^{-1} \hat V^{(w2b)}$ and $N^{-1} \hat V^{(w2)}$.
CMLE or oldBP denotes results from Cox's regression \texttt{coxph} with exact calculation or Breslow--Peto approximation in R package \texttt{survival}.}
\end{center} \vspace{-.2in}
\end{table}

To further compare estimators, we also conduct a simulation study.
For each simulation, a sample of size $n=200$ is generated as follows. The group variable $Z$ is generated as 1 or 2 with probability $.5$ each.
The event time $\tilde T$ is generated as Weibull with shape and scale parameters 2 and 1 or respectively 1 and 1 in the first or second group.
The censoring time $\tilde C$ is generated as 4 times Beta$(2,2)$ in the first group, and uniform$(0,4)$ in the second group.
Two sets of observed data $(Y,\delta)$ are then obtained, where $\delta = 1\{ \tilde T \le \tilde C\}$
and $Y$ is defined by discretizing $\tilde Y=\min(\tilde T, \tilde C)$ in intervals of length $.01$ or $.2$,
using the censored-late option discussed in the Supplement.
Models (\ref{eq:oddr-model2}) and (\ref{eq:pr-model2}) are fit with the time functions $x(t)=(1, x_1(t))^\T$, where $x_1(t)=1\{t > 1\}$.
Both models are in principle misspecified but can be considered an approximation to the underlying relative hazards over time,
which are plotted in the Supplement.

Table~\ref{tab:survival-sim} presents the results from 2000 repeated simulations.
There are similar patterns in these results as in the preceding discussion of Table~\ref{tab:VA}.
An additional confirmation from the simulations is that the model-robust variance estimator
(as well as the new model-based variance estimator) reasonably matches the Monte Carlo standard deviation for the BP estimator,
whereas the commonly reported model-based variance estimator on the row oldBP is upward biased.

\begin{table}
\caption{Comparison of estimates from simulations of survival data} \label{tab:survival-sim}
\small
\begin{center}
\begin{tabular*}{.85\textwidth}{ll rrrr l rrrr} \hline
     && $\hat\beta_0$  & SD       & bSE      & rSE     && $\hat\beta_1$  &  SD  &  bSE      & rSE \\ \hline
     && \multicolumn{9}{c}{Finely discretized data} \\
MH   && $-.2228$ & $.1932$  & $.1912$  & $.1928$ && $1.2158$ & $.4041$ & $.4035$ & $.3980$ \\
wMH  && $-.2182$ & $.1885$  & $.1885$  & $.1883$ && $1.2168$ & $.4045$ & $.4015$ & $.3986$ \\
CML  && $-.2184$ & $.1887$  & $.1885$  & NA      && $1.2164$ & $.4041$ & $.4023$ & NA \\
oldBP&& $-.2162$ & $.1868$  & $.1876$  & $.1866$ && $1.1979$  & $.3984$ & $.3993$ & $.3927$ \\
BP   && $-.2162$ & $.1868$  & $.1867$  & $.1866$ && $1.1979$  & $.3984$ & $.3964$ & $.3927$ \\  \cline{2-11}

     && \multicolumn{9}{c}{Coarsely discretized data} \\
MH   && $-.2388$ & $.2081$ & $.2102$  & $.2082$ && $1.3842$  & $.4545$ & $.4559$ & $.4506$ \\
wMH  && $-.2319$ & $.2015$ & $.2053$  & $.2018$ && $1.3912$  & $.4603$ & $.4563$ & $.4555$ \\
CML  && $-.2376$ & $.2065$ & $.2070$  & NA      && $1.3820$  & $.4537$ & $.4557$ & NA \\
oldBP&& $-.1949$ & $.1694$ & $.1877$  & $.1697$ && $1.0300$  & $.3474$ & $.3931$ & $.3421$ \\
BP   && $-.1949$ & $.1694$ & $.1723$  & $.1697$ && $1.0300$  & $.3474$ & $.3445$ & $.3421$ \\ \hline
\end{tabular*} \\[1ex]
\parbox{.85\textwidth}{\scriptsize Note: Point ($\hat\beta_0$ or $\hat\beta_1$) and SD are as defined as in Table~\ref{tab:survival-sim}.}
\end{center} \vspace{-.2in}
\end{table}

\vspace{.3in}
\centerline{\bf\Large References}

\begin{description}\addtolength{\itemsep}{-.1in}
\item Allison, P.D. (1982) Discrete-time methods for the analysis of event histories, {\em Sociological Methodology}, 13, 61--98.

\vspace{-.05in} \item Andersen, P.K., Borgan, O., Gill, R.D., and Keiding, N. (1993) {\em Statistical Models Based on Counting Processes}, New York: Springer.

\vspace{-.05in} \item Breslow, N.E. (1974) Covariance analysis of censored survival data, {\em Biometrics},  30, 89--100.

\vspace{-.05in} \item Breslow, N.E. (1976) Regression analysis of the log odds ratio: A method for retrospective studies, {\em Biometrics}, 32, 409--416.

\vspace{-.05in} \item Breslow, N.E. (1981) Odds ratio estimators when the data are sparse, {\em Biometrika}, 68, 73--84.

\vspace{-.05in} \item Breslow, N.E. and Cologne, J. (1986) Methods of estimation in log odds ratio regression models, {\em Biometrics}, 42, 949--954.

\vspace{-.05in} \item Breslow, N.E. and Day, N.E. (1980) {\em Statistical Methods in Cancer Research, Volume 1: The Analysis of Case-Control Studies}. Lyon: International Agency for Research on Cancer.


\vspace{-.05in} \item Buja, A., Berk, R., Brown, L., George, E., Pitkin, E., Traskin, M., Zhao, L., and Zhang, K. (2019) Models as approximations I:
Consequences illustrated with linear regression, {\em Statistical Science}, to appear.

\vspace{-.05in} \item Cochran, W.G. (1954) Some methods for strengthening the common $\chi^2$ tests, {\em Biometrics}, 10, 417--451.

\vspace{-.05in} \item Cox, D.R. (1972) Regression models and life tables (with discussion), {\em Journal of the Royal Statistical Society}, Ser.~B, 34, 187--220.

\vspace{-.05in} \item Cox, D.R. and Oaks, D.O. (1984) {\em Analysis of Survival Data}, London: Chapman \& Hall.

\vspace{-.05in} \item Davis, L.J. (1985) Generalization of the Mantel--Haenszel estimator to nonconstant odds ratios, {\em Biometrics}, 41, 487--495.

\vspace{-.05in} \item Efron, B. (1977) The efficiency of Cox's likelihood function for censored data, {\em  Journal of the American Statistical Association}, 72, 557--565.

\vspace{-.05in} \item Guilbaud, O. (1983)  On the large-sample distribution of the Mantel--Haenszel odds-ratio estimator, {\em Biometrics}, 39, 523--525.


\vspace{-.05in} \item Hauck, W.W. (1984) A comparative study of conditional maximum likelihood estimation of a common odds ratio, {\em Biometrics}, 40, 1117--1123.

\vspace{-.05in} \item Hauck, W.W., Anderson, S., Leahy, S.J.III (1982) Finite-Sample properties of some old and some new estimators of a common Odds ratio from multiple 2 × 2 tables,
{\em  Journal of the American Statistical Association}, 77, 145--152.

\vspace{-.05in} \item Kalbfleisch, J.D. and Prentice, R.L. (1980) {\em The Statistical Analysis of Failure Time Data}, New York: Wiley.

\vspace{-.05in} \item Kaplan, E.L. and Meier, P. (1958) Nonparametric estimation from incomplete observations, {\em Journal of the American Statistical Association}, 53, 457--481.

\vspace{-.05in} \item Kuritz, S.J.,  Landis, R., and Koch, G.G. (1988) A general overview of Mantel--Haenszel methods: Applications and recent developments,
{\em Annal Review of Public Health}, 9, 123--160.

\vspace{-.05in} \item Lin, D.Y. and Wei, L.J. (1989) The robust inference for the Cox proportional hazards model, {\em Journal of the American Statistical Association}, 84, 1074--1079.

\vspace{-.05in} \item Lindsay, B.G. (1980) Nuisance parameters, mixture models and the efficiency of partial likelihood estimators, {\em Philosophical Transactions of the Royal Society}, Series A, 296, 639--665.

\vspace{-.05in} \item Manski, C.F. (1988) {\em Analog Estimation Methods in Econometrics}. New York: Chapman \& Hall.

\vspace{-.05in} \item Mantel, N. and Haenszel, W.M. (1959)
Statistical aspects of the analysis of data from retrospective studies of disease, {\em Journal of the National Cancer Institute}, 22, 719--748.

\vspace{-.05in} \item McCullagh, P. and Nelder, J. (1989) {\em Generalized Linear Models}, 2nd ed, New York: Chapman \& Hall.

\vspace{-.05in} \item Peto, R. (1972) Contribution to the discussion of Cox (1972): Regression models and life tables, {\em Journal of the Royal Statistical Society}, Ser.~B, 34, 205--207.

\vspace{-.05in} \item Prentice, R.L. and Gloeckler, L.A. (1978) Regression analysis of grouped survival data with application to breast cancer data, {\em Biometrics}, 34, 57--67.

\vspace{-.05in} \item Robins, J.M., Breslow, N.E., Greenland, S. (1986) Estimators of the Mantel--Haenszel variance consistent in both
sparse data and large strata limiting models, {\em Biometrics}, 42, 311--324.

\vspace{-.05in} \item Tan, Z. (2006) A distributional approach for causal inference using propensity scores, {\em Journal of the American Statistical Association}, 101, 1619--1637.

\vspace{-.05in} \item Thompson, W.A.Jr. (1977) On the treatment of grouped observations in life studies, {\em Biometrics}, 33, 463--470.

\vspace{-.05in} \item Therneau, T.M. (2015) {\em A Package for Survival Analysis}, version 2.38.


\vspace{-.05in} \item White, H. (1982) Maximum likelihood estimation of misspecified models, {\em Econometrica}, 50, 1–25.

\vspace{-.05in} \item Zelen, M. (1971) The analysis of several 2 x 2 contingency tables, {\em Biometrika}, 58, 129--137.
\end{description}

\clearpage

\setcounter{page}{1}

\setcounter{section}{0}
\setcounter{equation}{0}

\setcounter{figure}{0}
\setcounter{table}{0}

\renewcommand{\theequation}{S\arabic{equation}}
\renewcommand{\thesection}{\Roman{section}}

\renewcommand\thefigure{S\arabic{figure}}
\renewcommand\thetable{S\arabic{table}}

\begin{center}
{\Large Supplementary Material for}

{\Large ``Analysis of odds, probability, and hazard ratios: From 2 by 2 tables to two-sample survival data"}

\vspace{.1in} {\large Zhiqiang Tan}
\end{center}
\vspace{.1in}

\section{Variance estimation for Mantel--Haenszel}

For completeness, we discuss various model-based variance estimators for the Mantel--Haenszel estimator $\hat\beta^{(0)}$ under the assumption of common odds ratios, $\psi_1^* =\ldots=\psi_J^* = \exp(\beta^*)$,
i.e., model (\ref{eq:oddr-model}) is valid with $x_1=\ldots=x_J=1$.

The (non-symmetrized) variance estimator for $\hat\beta^{(0)}$ in Robins et al.~(1986) is
\begin{align*}
\hat \Sigma^{(b)} (\hat\beta^{(0)}) =\frac{ \sum_{j=1}^J   \rho_j^{(0)}   \left\{ \hat B_j \frac{ N_{2j}}{N_j}  \left( \frac{\hat p_{22j}}{\hat\psi}  + \hat p_{21j} \right)
 +  \hat A_j \frac{ N_{1j}}{N_j} \left( \frac{\hat p_{11j}}{\hat\psi^2} +  \frac{\hat p_{12j}}{\hat\psi } \right) \right\} } { \left(\sum_{j=1}^J \rho_j^{(0)}\hat B_j \right)^2 },
\end{align*}
where $\hat\psi  =\exp( \hat\beta^{(0)})$, $\hat A_j = \hat p_{11j} \hat p_{22j}$ and $\hat B_j = \hat p_{12j} \hat p_{21j}$.
By exchange of the response values, i.e., $(\hat p_{11j}, \hat p_{21j})$ with $(\hat p_{12j}, \hat p_{22j})$, the preceding estimator can be shown to be
\begin{align*}
\tilde \Sigma^{(b)} (\hat\beta^{(0)}) =\frac{ \sum_{j=1}^J   \rho_j^{(0)}   \left\{ \hat A_j \frac{ N_{2j}}{N_j}  \left( \frac{\hat p_{21j}}{\hat\psi}  + \frac{\hat p_{22j}}{\hat\psi^2} \right)
 +  \hat B_j \frac{ N_{1j}}{N_j} \left( \hat p_{12j}  +  \frac{\hat p_{11j}}{\hat\psi } \right) \right\} } { \left(\sum_{j=1}^J \rho_j^{(0)}\hat B_j \right)^2 },
\end{align*}
using the fact that $\hat\psi = \sum_{j=1}^J   \rho_j^{(0)} \hat A_j / \sum_{j=1}^J   \rho_j^{(0)} \hat B_j$.
The symmetrized variance estimator for $\hat\beta^{(0)}$ in Robins et al.~(1986) is $\{\hat \Sigma^{(b)} (\hat\beta^{(0)}) + \tilde \Sigma^{(b)} (\hat\beta^{(0)}) \}/2$.

The variance estimator for $\hat\beta^{(0)}$ in Flander (1985) is
\begin{align*}
\hat \Sigma^{(b2)} (\hat\beta^{(0)}) =\frac{ \sum_{j=1}^J   \rho_j^{(0)} \hat B_j \frac{ 1}{N_j}  \left( \frac{n_{11j}+n_{22j}+1}{\hat\psi}  + n_{12j} + n_{21j}-1 \right)
 } { \left(\sum_{j=1}^J \rho_j^{(0)}\hat B_j \right)^2 },
\end{align*}
which can be rewritten as
\begin{align*}
\hat \Sigma^{(b2)} (\hat\beta^{(0)}) =\frac{ \sum_{j=1}^J   \rho_j^{(0)} \hat B_j  \left\{ \frac{ N_{2j}}{N_j}  \left( \frac{\hat p_{22j}}{\hat\psi}  + \hat p_{21j} \right)
 + \frac{ N_{1j}}{N_j} \left( \hat p_{12j}  +  \frac{\hat p_{11j} }{\hat\psi } \right) + \frac{\hat\psi^{-1} -1}{N_j} \right\} } { \left(\sum_{j=1}^J \rho_j^{(0)}\hat B_j \right)^2 },
\end{align*}
The estimator obtained through exchange of the response values is then
\begin{align*}
\tilde \Sigma^{(b2)} (\hat\beta^{(0)}) =\frac{ \sum_{j=1}^J   \rho_j^{(0)}  \hat A_j  \left\{ \frac{ N_{2j}}{N_j}  \left( \frac{\hat p_{21j}}{\hat\psi}  + \frac{\hat p_{22j}}{\hat\psi^2} \right)
 +  \frac{ N_{1j}}{N_j} \left( \frac{\hat p_{11j}}{\hat\psi^2} +  \frac{\hat p_{12j}}{\hat\psi } \right)+ \frac{\hat\psi^{-1} - \hat\psi^{-2}}{N_j}   \right\} } { \left(\sum_{j=1}^J \rho_j^{(0)}\hat B_j \right)^2 } .
\end{align*}
The symmetrized version,  $\{\hat \Sigma^{(b2)} (\hat\beta^{(0)}) + \tilde \Sigma^{(b2)} (\hat\beta^{(0)}) \}/2$, can be shown to be
$\{\hat \Sigma^{(b)} (\hat\beta^{(0)}) + \tilde \Sigma^{(b)} (\hat\beta^{(0)}) \}/2$ plus
$(\hat\psi^{-1} - \hat\psi^{-2}) \{\sum_{j=1}^J   \rho_j^{(0)}  (\hat A_j - \hat\psi \hat B_j) /N_j \} / (\sum_{j=1}^J \rho_j^{(0)}\hat B_j)^2 /2$.

Similarly as those in Robins et al.~(1986), the estimators $\hat \Sigma^{(b2)} (\hat\beta^{(0)})$, $\tilde \Sigma^{(b2)} (\hat\beta^{(0)})$, and the symmetrized version can be shown to be consistent variance estimators for $\hat\beta^{(0)}$
in both Settings I and II. In fact, the variance estimator $\hat \Sigma^{(b2)} (\hat\beta^{(0)})$ corresponds to use of the following estimator of $\sigma_j(\beta)$
instead of $\hat \sigma^{(b)}_j(\beta)$:
\begin{align*}
\hat \sigma^{(b2)}_j(\beta) & =  \psi_j(\beta) \hat p_{12j}\hat p_{21j} \left\{ \frac{\hat p_{22j} + \psi_j(\beta) \hat p_{21j}}{ N_{1j} }
 +  \frac{\hat p_{11j} + \psi_j(\beta) \hat p_{12j} }{N_{2j}} + \frac{1-\psi_j(\beta)}{N_{1j} N_{2j}} \right\}.
\end{align*}
Both $\hat \sigma^{(b)}_j(\beta^*)$ and $\hat \sigma^{(b2)}_j(\beta^*) $
can be shown to be unbiased for  $ \sigma^{(b)}_j(\beta^*)$, that is,
$E\{ \hat \sigma^{(b2)}_j(\beta^*) \} = E\{\hat \sigma^{(b)}_j(\beta^*) \} $, which is equivalent to
\begin{align*}
E[ \hat p_{11j}\hat p_{22j} \{\hat p_{11j} + \psi_j(\beta^*) \hat p_{12j} \}] & =
 E[ \psi_j(\beta^*) \hat p_{12j}\hat p_{21j} \{\hat p_{11j} + \psi_j(\beta^*) \hat p_{12j} +  (1-\psi_j(\beta^*))/N_{1j} \} ].
\end{align*}
This can be verified with the following calculation:
\begin{align*}
& E[ \hat p_{11j}\hat p_{22j} \{\hat p_{11j} + \psi_j(\beta^*) \hat p_{12j} \}] \\
& = p_{11j} p_{22j} \{ p_{11j} + \psi_j(\beta^*) p_{12j} \} + \frac{p_{11j} p_{12j}}{N_{1j}} p_{22j} - \frac{p_{11j} p_{12j}}{N_{1j}} p_{22j}  \psi_j(\beta^*) , \\
& E[ \psi_j(\beta^*) \hat p_{12j}\hat p_{21j} \{\hat p_{11j} + \psi_j(\beta^*) \hat p_{12j}\} ] \\
& = \psi_j(\beta^*) p_{12j} p_{21j} \{ p_{11j} + \psi_j(\beta^*) p_{12j}\} - \frac{p_{11j} p_{12j}}{N_{1j}} p_{21j} \psi_j(\beta^*)+ \frac{p_{11j} p_{12j}}{N_{1j}} p_{21j}\psi_j^2(\beta^*),
\end{align*}
using the fact that $E ( \hat p _{11j}^2 ) = p_{11j}^2 + p_{11j} (1-p_{11j}) / N_{1j}$, $E ( \hat p _{11j} \hat p_{12j} ) = p_{11j} p_{12j} - p_{11j} (1-p_{11j}) / N_{1j}$, etc.

\section{Grouped survival data}

Consider the situation where survival data are collected by grouping continuous or fine-scaled measurements, but such detailed data are not available.
Let $0 = t_0 < t_1 < \cdots < t_J$ be fixed inspection points.
A typical procedure for data collection is as follows (Kaplan \& Meier 1958): for $j=1, \ldots, J$,
examine a cohort of individuals of age $t_j$ and determine whether each individual survived the interval, experienced an event, or was lost from the study during the interval.
By convention, a censored time in the continuous or fine scale at $t_j$ means an individual being seen event-free at $t_j$ but not after. Hence
each censored time found in the $j$th inspection falls in $[t_{j-1}, t_j)$, although its actual location is unknown. A censored time at $t_j$  would be counted in the next inspection.

We discuss how the variables $(Y,\delta)$ can be defined in a discrete scale for such grouped data.
For an event observed in $(t_{j-1},t_j]$, it is natural to set $Y=t_j$ and $\delta=1$.
A complication is that there are different options for treating censored times, corresponding to different assumptions (Kaplan \& Meier 1958; Thompson 1977).
One extreme (referred to as censored-early) is to assume that each censored time in $[t_{j-1}, t_j)$ occur at $t_{j-1}$ and hence
set $Y=t_{j-1}$ and $\delta=0$.
The other extreme (referred to as censored-late) is to assume that each censored time in $[t_{j-1}, t_j)$ occur immediately before $t_j$ and hence
set $Y=t_j$ and $\delta=0$.
Effectively, the censored-early option excludes the censored observations in $[t_{j-1}, t_j)$, whereas
the censored-late option includes those observations, in the risk set associated with $t_j$ (Cox 1972).
In general, each of the two options for defining the variables $(Y,\delta)$ is a biased approximation when censored times may occur strictly inside the intervals.
On the other hand, if censored times can occur only at the discrete inspection points, then the first option is appropriate whereas
the second option would encode a censored time at $t_{j-1}$ incorrectly as $t_j$.

The preceding discussion deals with the definitions of $(Y,\delta)$. To use the standard framework for survival analysis, we need also
to discuss the definitions of the full-data variables $(T,C)$ and the conditional independent censoring assumption
after reduction of continuous data to grouped data. For simplicity, the covariate is conditioned on. Let $(\tilde Y, \delta)$ be the censored data,
associated with the event and censoring times $(\tilde T, \tilde C)$ in the continuous or fine scale
such that $\tilde Y = \min(\tilde T,\tilde C)$ and $ \delta = 1\{\tilde T \le \tilde C\}$, the same as in the grouped data.
The censored-early option  for defining $(Y, \delta)$ is as follows:
\begin{itemize} \addtolength{\itemsep}{-.1in}
\item[($i$)] set $Y= t_j$ and $\delta =1$ if $\tilde Y \in (t_{j-1}, t_j]$ and $\tilde Y = \tilde T \le \tilde C$, or
\item[($ii$)] set $Y= t_{j-1}$ and $\delta =0$ if $\tilde Y \in [t_{j-1}, t_j)$ and $\tilde T > \tilde C = \tilde Y$.
\end{itemize}
To ensure the relationship $Y = \min(T,C)$ and $\delta=1\{Z \le C\}$, the corresponding variables $(T,C)$ can be defined as follows, with $t_k \ge t_j$ in ($iii$) and ($iv$):
\begin{itemize} \addtolength{\itemsep}{-.1in}
\item[($iii$)] set $T= t_j$ and $C =t_k$ for $\tilde C \in [t_{k-1},t_k)$ if $\tilde Y \in (t_{j-1}, t_j]$ and $\tilde Y = \tilde T \le \tilde C$, or
\item[($iv$)] set $C= t_{j-1}$ and $T =t_k$ for $\tilde T \in (t_{k-1},t_k]$ if $\tilde Y \in [t_{j-1}, t_j)$ and $\tilde T > \tilde C = \tilde Y$.
\end{itemize}
Although $T$ is defined deterministically from $\tilde T$, the definition of $C$ depends on whether $\tilde T \le \tilde C$ or $\tilde T > \tilde C$.
Hence $T$ and $C$ in general fail to be independent, even though $\tilde T $ and $\tilde C$ are independent.
This issue can also been seen from another angle. The variable $C$ can be redefined deterministically as $C =t_{j-1}$ for $\tilde C \in [t_{j-1}, t_j)$.
Then $Y = \min(T,C)$ and  $\delta =1\{T \le C\}$ would in general hold only under the restriction that $\tilde C \ge t_j$ if $\tilde Y \in (t_{j-1}, t_j]$ and $\tilde Y = \tilde T \le \tilde C$.
But this restriction would contradict the independence of $\tilde T$ and $\tilde C$, unless $\tilde C$ is discrete, taking values in $\{t_0,t_1,\ldots,t_J\}$.

A concrete consequence of the violation of the independence of $T$ and $C$ is that the event probability calculated within a risk set in general differs from the
conditional probability of the event time.  In fact, it can be shown from the definitions in ($i$) and ($ii$), regardless of ($iii$) and ($iv$), that
\begin{align}
& P ( Y = t_j, \delta =1 | Y \ge t_j ) = \frac{P( \tilde T \in (t_{j-1} ,t_j], \tilde T \le \tilde C ) }{ P (\tilde T > t_{j-1}, \tilde T \le \tilde C) + P(\tilde C \ge t_j, \tilde T > \tilde C)} \nonumber \\
& = \frac{ \pi_j+ P(\tilde T \in (t_{j-1},t_j], \tilde C \ge t_j) }
{\pi_j +  P(\tilde T  >t_{j-1}, \tilde C \ge t_j)  }  \label{eq:approx-prob} \\
& \ge P (\tilde T \in (t_{j-1},t_j] | \tilde T > t_{j-1}),  \nonumber
\end{align}
where $\pi_j = P ( \tilde T \in (t_{j-1}, t_j), \tilde C \in (t_{j-1},t_j), \tilde T \le \tilde C)$.
The probability $P ( Y = t_j, \delta =1 | Y \ge t_j )$ identified from the risk set
is in general an over-approximation of the desired conditional probability $ P (\tilde T \in (t_{j-1},t_j] | \tilde T > t_{j-1})$.
The bias incurred can be negligible if the intervals are small.
As an exception, the two probabilities coincide in the case of $\pi_j=0$, which is satisfied when $\tilde C$ is discrete, taking values in $\{t_0,t_1,\ldots,t_J\}$, as previously noticed.
Nevertheless, the methods in Section \ref{sec:survival} can be seen to provide valid inference about
odds and probability ratios defined from probabilities in (\ref{eq:approx-prob}).

There exist similar issues with the censored-late option discussed above for defining $(Y, \delta)$, which amounts to ($i$) and ($ii$) replaced by
\begin{itemize} \addtolength{\itemsep}{-.1in}
\item[($ii^\prime$)] set $Y= t_j$ and $\delta =0$ if $\tilde Y \in [t_{j-1}, t_j)$ and $\tilde T > \tilde C = \tilde Y$.
\end{itemize}
To ensure the relationship $Y = \min(T,C)$ and $\delta=1\{Z \le C\}$,  the corresponding variables $(T,C)$ can be defined as ($iii$) and ($iv$) replaced by
\begin{itemize} \addtolength{\itemsep}{-.1in}
\item[($iv^\prime$)] set $C= t_j$ and $T =t_{k+1}$ for $\tilde T \in (t_{k-1},t_k]$ if $\tilde Y \in [t_{j-1}, t_j)$ and $\tilde T > \tilde C = \tilde Y$.
\end{itemize}
Then $T$ and $C$ in general fail to be independent, even though $\tilde T $ and $\tilde C$ are independent.
From the definitions in ($i$) and ($ii^\prime$), regardless of ($iii$) and ($iv^\prime$), it can be shown that
\begin{align}
& P ( Y = t_j, \delta =1 | Y \ge t_j ) = \frac{P( \tilde T \in (t_{j-1} ,t_j], \tilde T \le \tilde C ) }{ P (\tilde T > t_{j-1}, \tilde T \le \tilde C) + P(\tilde C \ge t_{j-1}, \tilde T > \tilde C)} \nonumber \\
& = \frac{  P(\tilde T \in (t_{j-1},t_j], \tilde C \ge t_{j-1}) - \pi^\prime_j }
{ P(\tilde T  >t_{j-1}, \tilde C \ge t_{j-1})  }  \label{eq:approx-prob2} \\
& \le P (\tilde T \in (t_{j-1},t_j] | \tilde T > t_{j-1}),  \nonumber
\end{align}
where $\pi_j^\prime = P ( \tilde T \in (t_{j-1}, t_j], \tilde C \in [t_{j-1},t_j), \tilde T > \tilde C)$.
The probability $P ( Y = t_j, \delta =1 | Y \ge t_j )$ identified from the risk set
is in general an under-approximation of the desired conditional probability $ P (\tilde T \in (t_{j-1},t_j] | \tilde T > t_{j-1})$.
The two probabilities remain different even when $\tilde C$ is discrete, taking values in $\{t_0,t_1,\ldots,t_J\}$, in agreement with our earlier comment.
Nevertheless, the methods in Section \ref{sec:survival} can be seen to provide valid inference about
odds and probability ratios defined from probabilities in (\ref{eq:approx-prob2}).

\section{Proofs}

{\noindent \textbf{Proof of Proposition~\ref{pro:optimal-EE}.}}

(i) For simplicity, write $G_\rho=G_\rho(\beta^*)$ and $H_\rho=H_\rho(\beta^*)$.
For the claimed optimal choice $\rho^\dag_j(\beta)$, it is easily shown that $G_{\rho^\dag}=H_{\rho^\dag}$ and hence the asymptotic variance $H_{\rho^\dag}^{-1} G_{\rho^\dag} H_{\rho^\dag}^{-1}$ reduces to $G_{\rho^\dag}^{-1}$.
For another choice $\rho_j (\beta)$, we need to show that
$H_{\rho}^{-1} G_{\rho} H_{\rho}^{-1} \ge G_{\rho^\dag}^{-1}$
or equivalently $G_{\rho} \ge H_{\rho} G_{\rho^\dag}^{-1} H_{\rho} $.
This follows from the matrix version of the Cauchy--Schwartz inequality (e.g., Proof of Theorem 1, Tan 2004)
\begin{align*}
\var  \left\{ \tau_{\rho}(\beta^*) \right\} \ge H_{\rho}
\var^{-1}  \left\{ \tau_{\rho^\dag}(\beta^*) \right\} H_{\rho}^\T,
\end{align*}
where $G_\rho = N_\bullet^{-1} \var\{  \tau_{\rho}(\beta^*) \}$ and
$H_{\rho} = N_\bullet^{-1} \cov\{ \tau_{\rho}(\beta^*), \tau_{\rho^\dag}(\beta^*) \}$ by direct calculation.

(ii) We show that $\rho_j^\dag(\beta^*)$ can be approximated up to negligible terms by $\rho_j^\ddag(\beta^*)$.
In Setting I, $\var ( \hat p_{11j} \hat p_{22j} - \psi_j^* \hat p_{12j} \hat p_{21j} )$ can be approximated up to terms of order $o(N_j^{-1})$ by
$\var \{ (\hat p_{11j} - p_{11j}) ( p_{22j} + \psi^*_j p_{21j}) + (\hat p_{21j}- p_{21j}) (p_{11j} + \psi^*_j p_{12j}) \} $. By the independence of
$\hat p_{11j}$ and $\hat p_{21j}$, this variance can be calculated as follows:
\begin{align*}
& \frac{ p_{11j}p_{12j}}{N_{1j}} ( p_{22j} + \psi^*_j p_{21j})^2 + \frac{ p_{21j} p_{22j}}{N_{2j}} (p_{11j} + \psi^*_j p_{12j})\\
& =   \frac{\psi_j^* p_{12j} p_{21j}}{ N_{1j} N_{2j}} \left\{ N_{2j}( p_{22j} + \psi^*_j p_{21j}) + N_{1j} (p_{11j} + \psi^*_j p_{12j})  \right\},
\end{align*}
using the fact that $p_{22j} + \psi^*_j p_{21j} = p_{22j}/p_{12j}$ and $p_{11j} + \psi^*_j p_{12j} = p_{11j} /p_{21j}$.
Substituting this approximation into $\rho_j^\dag(\beta^*)$ yields $\rho_j^\ddag(\beta^*)$ for $j=1,\ldots,J$.
{\hfill $\Box$ \vspace{.1in}}

{\noindent \textbf{Proof of Proposition~\ref{pro:var-est}.}}

(i) Denote $q_{1j} = \hat p_{11j}\hat p_{12j} /(N_{1j}-1)$ and $q_{2j} = \hat p_{21j} \hat p_{22j} /(N_{2j}-1)$.
First, we show that  if $n_{12j} \ge 1$ and $n_{22j}\ge 1$ or equivalently $\hat p_{12j} \ge 1/N_{1j}$ and $\hat p_{22j}\ge 1/N_{2j}$, then
$h(u) \ge 0$ for $u > 0$, where
\begin{align*}
h(u) &=  q_{1j}(\hat p_{22j} + u \hat p_{21j} )^2 +  q_{2j} (\hat p_{11j} + u\hat p_{12j} )^2
- (u-1)^2q_{1j} q_{2j} .
\end{align*}
The quadratic coefficient of $h(u)$ is $q_{1j} \hat p_{21j}^2 + q_{2j} \hat p_{12j}^2 - q_{1j} q_{2j} \ge q_{2j} ( \hat p_{12j}^2 - q_{1j}) \ge 0$ because
$\hat p_{12j} \ge \hat p_{11j}/(N_{1j}-1)$ or $\hat p_{12j} \ge 1/N_{1j}$.
Hence the derivative of $h(u)$, denoted as $h^\prime(u)$, is non-decreasing. But $h^\prime(0) = 2q_{1j} \hat p_{21j}\hat p_{22j} + 2 q_{2j} \hat p_{11j} \hat p_{12j} + 2 q_{1j}q_{2j} \ge 0$.
Hence $h^\prime(u) \ge 0$ and $h(u)$ is non-decreasing for $u \ge 0$. The desired inequality then follows because $h(0) = q_{1j} \hat p_{22j}^2 +  q_{2j} \hat p_{11j}^2 - q_{1j} q_{2j}
\ge q_{1j} ( \hat p_{22j}^2 - q_{2j}) \ge 0$ because $\hat p_{22j} \ge \hat p_{21j}/(N_{2j}-1)$ or $\hat p_{22j} \ge 1/N_{2j}$.
Second, if $n_{11j} \ge 1$ and $n_{21j}\ge 1$ or equivalently $\hat p_{11j} \ge 1/N_{1j}$ and $\hat p_{21j}\ge 1/N_{2j}$, then
the preceding argument can be applied to show that for any $u >0$,
\begin{align*}
&  q_{1j}(u^{-1} \hat p_{22j} + \hat p_{21j} )^2 +  q_{2j} (u^{-1} \hat p_{11j} + \hat p_{12j} )^2
- (u^{-1} -1)^2q_{1j} q_{2j}\ge 0 ,
\end{align*}
by exchanging $(\hat p_{11j},\hat p_{21j})$ with $(\hat p_{12j},\hat p_{22j})$ and $u$ with $u^{-1}$.
Finally, $\hat \sigma_j(\beta)=0$ in the remaining cases where either $n_{12j} = 0$ and $n_{21j}= 0$ or $n_{11j} = 0$ and $n_{22j}= 0$.

(ii) First, we calculate $\sigma_j(\beta)$. In fact, consider the following decomposition:
\begin{align}
& \hat p_{11j} \hat p_{22j} - \psi_j (\beta) \hat p_{12j} \hat p_{21j} - \{ p_{11j} p_{22j} - \psi_j (\beta) p_{12j} p_{21j} \} \nonumber \\
& = (\hat p_{11j} - p_{11j}) \{ \hat p_{22j} + \psi_j(\beta) \hat p_{21j}\} -  (\hat p_{21j} - p_{21j}) \{ p_{11j} + \psi_j(\beta)  p_{12j} \} \nonumber \\
& = (\hat p_{11j} - p_{11j}) \{p_{22j} + \psi_j(\beta) p_{21j} \} -  (\hat p_{21j} - p_{21j}) \{  p_{11j} + \psi_j(\beta)  p_{12j} \} \nonumber \\
& \quad + \{\psi_j(\beta)-1\} (\hat p_{11j} - p_{11j})  (\hat p_{21j} - p_{21j}) \label{eq:decomp-MH}
\end{align}
By the independence of $\hat p_{11j}$ and $\hat p_{21j}$, we have
\begin{align*}
\sigma_j(\beta) & = \frac{p_{11j} p_{12j}}{N_{1j}} \{p_{22j} + \psi_j(\beta) p_{21j} \}^2  + \frac{p_{21j} p_{22j}}{N_{2j}} \{ p_{11j} + \psi_j(\beta)  p_{12j} \}^2 \\
& \quad + \{\psi_j(\beta)-1\}^2  \frac{p_{11j} p_{12j}}{N_{1j}}   \frac{p_{21j} p_{22j}}{N_{2j}} .
\end{align*}
Noting $E ( \hat p _{21j}^2 ) = p_{21j}^2 + p_{22j} (1-p_{21j}) / N_{2j}$, we calculate
\begin{align*}
& E [ \{\hat p_{22j} + \psi_j(\beta) \hat p_{21j} \}^2 ] = E ( [1 + \{\psi_j(\beta)-1\} \hat p_{21j} ]^2 ) \\
& = [1 + \{\psi_j(\beta)-1\} p_{21j} ]^2 + \{\psi_j(\beta)-1\}^2 p_{21j} (1-p_{21j}) / N_{2j} .
\end{align*}
Similarly, noting $E ( \hat p _{11j}^2 ) = p_{11j}^2 + p_{11j} (1-p_{11j}) / N_{1j}$, we calculate
\begin{align*}
& E[ \{\hat p_{11j} + \psi_j(\beta) \hat p_{12j} \}^2 ] = E ( [\psi_j(\beta) + \{1-\psi_j(\beta)\}  \hat p_{11j} ]^2 ) \\
& = E ( [\psi_j(\beta) + \{1-\psi_j(\beta)\}  p_{11j} ]^2 ) + \{\psi_j(\beta)-1\}^2 p_{11j} (1-p_{11j}) / N_{1j}.
\end{align*}
By the independence of $\hat p_{11j}$ and $\hat p_{12j}$, direct calculation using the preceding results
and the fact that $\hat p_{11j}\hat p_{12j} / (N_{1j}-1)$ is unbiased for $p_{11j} p_{12j} /N_{1j}$ and
$\hat p_{21j} \hat p_{22j} /(N_{2j}-1)$ is unbiased for  $p_{21j} p_{22j} /N_{2j}$ yields $E\{\hat \sigma_j(\beta) \} = \sigma_j(\beta)$.

(iii) The result follows from (ii) and similar arguments as in Robins et al.~(1986).
{\hfill $\Box$ \vspace{.1in}}

{\noindent \textbf{Proof of Proposition~\ref{pro:pr-optimal-EE}.}}

The choice $q^\dag_j(\gamma)$ can be obtained similarly as Proposition~\ref{pro:optimal-EE}(i). For optimality of $q^\ddag_j(\gamma)$, we show that  $q_j^\ddag (\gamma^*) =q_j^\dag (\gamma^*)$. In fact,
$\var \{ \hat p_{11j} - \phi_j (\gamma^*) \hat p_{21j} \}$ can be calculated as
\begin{align*}
& \frac{ p_{11j} p_{12j}}{N_{1j}} + \phi_j^{*2} \frac{ p_{21j} p_{22j}}{N_{2j}}  = \frac{\phi_j^* p_{21j}}{N_{1j} N_{2j}}  ( N_{2j}  p_{12j} + N_{1j} \phi_j^* p_{22j} ),
\end{align*}
using the fact that $p_{11j} = \phi_j^*  p_{21j}$. Substituting this approximation into $q^\dag_j(\gamma^*)$ yields $q^\ddag_j(\gamma^*)$ for $j=1,\ldots,J$.
{\hfill $\Box$ \vspace{.1in}}

{\noindent \textbf{Proof of Proposition~\ref{pro:pr-var-est}.}}

By the independence of $\hat p_{11j}$ and $\hat p_{21j}$, $v_j(\gamma)$ is $\var \{ \hat p_{11j} - \phi_j (\gamma) \hat p_{21j} \} =p_{11j} p_{12j} /N_{1j} +  \phi_j^2(\gamma) p_{21j} p_{22j} /N_{2j}$.
Result (i) follows directly because $\hat p_{11j}\hat p_{12j} / (N_{1j}-1)$ is unbiased for $p_{11j} p_{12j} /N_{1j}$ and
$\hat p_{21j} \hat p_{22j} /(N_{2j}-1)$ is unbiased for  $p_{21j} p_{22j} /N_{2j}$.
Result (ii) follows because $E (  \phi_j^* \hat p_{12j} \hat p_{21j} ) = p_{12j} p_{11j}$ and
$E ( \hat p_{11j}  \hat p_{22j}) = \phi_j^* p_{21j} p_{22j}$.
{\hfill $\Box$ \vspace{.1in}}

{\noindent \textbf{Proof of Proposition~\ref{pro:pr-asym}.}}

Both (i) and (ii) follow from similar arguments as in Lin \& Wei (1989). In fact, it can be shown that $\hat\gamma^{(w)}$ converges in probability to $\bar \gamma^{(w2)}$ solving (\ref{eq:pr-target}),
and $N^{1/2} ( \hat\gamma^{(w)} - \bar \gamma^{(w2)})$ converges in distribution to $\N(0, V^{(w2)})$, with
$ V^{(w2)}$ in the stated sandwich form, where $D^{(w2)}(\gamma)$ is defined in Proposition~\ref{pro:pr-asym}, but $C^{(w2)}(\gamma)$ is defined such that
\begin{align*}
\frac{1}{N^{1/2}} \sum_{j=1}^J \frac{N_{2j} n_{11j} - \phi_j N_{1j} n_{21j}} { N_{1j} \phi_j + N_{2j}} x_j \to \N(0, C^{(w2)}(\gamma) ).
\end{align*}
Here and subsequently, $\phi_j=\phi_j(\gamma)$ and $\gamma$ is evaluated at $\bar\gamma^{(w2)}$. To complete the proof, it suffices to show that
\begin{align*}
& \frac{1}{N} \sum_{j=1}^J \frac{N_{2j} n_{11j} - \phi_j N_{1j} n_{21j}} { N_{1j} \phi_j + N_{2j}} x_j = \frac{1}{N} \sum_{i=1}^N  \sum_{j=1}^J  h_j(Y_i,\delta_i,Z_i;  \gamma ) x_j + o_p(N^{-1/2}).
\end{align*}
%
Consider the decomposition using (\ref{eq:pr-target}),
\begin{align*}
&\frac{1}{N} \sum_{j=1}^J \frac{N_{2j} n_{11j} - \phi_j N_{1j} n_{21j}} { N_{1j} \phi_j + N_{2j}} x_j \\
&= \frac{1}{N} \sum_{j=1}^J \frac{N_{2j} (n_{11j} - p_{11j} N_{1j} )  - \phi_j N_{1j} (n_{21j} - p_{21j} N_{2j} )} { N_{1j} \phi_j + N_{2j}} x_j \\
& \quad + \sum_{j=1}^J (p_{11j} - \phi_j p_{21j}) \left( \frac{N_{1j}N_{2j} /N}{N_{1j} \phi_j + N_{2j}} - \frac{P_{1j} P_{2j}}{ P_{1j} \phi_j + P_{2j}} \right) x_j.
\end{align*}
The first term can be shown to be
\begin{align*}
& \frac{1}{N} \sum_{j=1}^J \frac{N_{2j} (n_{11j} - p_{11j} N_{1j} )  - \phi_j N_{1j} (n_{21j} - p_{21j} N_{2j} )} { N_{1j} \phi_j + N_{2j}} x_j \\
& = \frac{1}{N} \sum_{j=1}^J \frac{P_{2j} (n_{11j} - p_{11j} N_{1j} )  - \phi_j P_{1j} (n_{21j} - p_{21j} N_{2j} )} { P_{1j} \phi_j + P_{2j}} x_j + o_p(N^{-1/2})  .
\end{align*}
The second term can be shown to be
\begin{align*}
& \sum_{j=1}^J (p_{11j} - \phi_j p_{21j}) \left( \frac{N_{1j}N_{2j} /N}{N_{1j} \phi_j + N_{2j}} - \frac{P_{1j} P_{2j}}{ P_{1j} \phi_j + P_{2j}} \right) x_j \\
& = \sum_{j=1}^J (p_{11j} - \phi_j p_{21j}) \frac{(\frac{N_{1j}}{N}-P_{1j}) P_{2j}^2 + ( \frac{N_{2j}}{N} - P_{2j})\phi_j P_{1j}^2 }{ (P_{1j} \phi_j + P_{2j})^2} x_j + o_p(N^{-1/2})\\
& = \sum_{j=1}^J (p_{11j} - \phi_j p_{21j}) \frac{ \frac{N_{1j}}{N} P_{2j}^2 +  \frac{N_{2j}}{N}  \phi_j P_{1j}^2 }{ (P_{1j} \phi_j + P_{2j})^2} x_j + o_p(N^{-1/2}),
\end{align*}
where the last step follows because by (\ref{eq:pr-target}),
\begin{align*}
& \sum_{j=1}^J (p_{11j} - \phi_j p_{21j}) \frac{ P_{1j}  P_{2j}^2 +  P_{2j} \phi_j P_{1j}^2 }{ (P_{1j} \phi_j + P_{2j})^2} x_j
 = \sum_{j=1}^J   (p_{11j} - \phi_j p_{21j}) \frac{P_{1j} P_{2j}}{ P_{1j} \phi_j + P_{2j}} x_j =0 .
\end{align*}
The desired expansion follows by collecting the preceding results.

As an auxiliary result, we show that $\hat h_j(Y,\delta,Z; \gamma)$ coincides with $\hat h_j^{\mbox{\tiny LW}}(Y,\delta,Z; \gamma)$ in Remark~\ref{rem:lin-wei}.
By definition, $\hat h_j(Y,\delta,Z; \gamma)$ can be calculated as
\begin{align*}
\hat h_j(Y,\delta,Z; \gamma) &= \frac{N_{2j}}{ N_{1j} \phi_j(\gamma) + N_{2j}} ( I_{11j} - \hat p_{11j} I_{1j} ) - \frac{N_{1j}\phi_j(\gamma)}{ N_{1j} \phi_j(\gamma) + N_{2j}} ( I_{21j} - \hat p_{21j} I_{2j} ) \\
& \quad + \frac{\hat p_{11j} - \phi_j(\gamma) \hat p_{21j}}{ \{N_{1j} \phi_j(\gamma) + N_{2j} \}^2} \{N_{2j}^2 I_{1j} + \phi_j(\gamma) N_{1j}^2 I_{2j} \} \\
& =  \frac{N_{2j}}{ N_{1j} \phi_j(\gamma) + N_{2j}}  I_{11j}  - \frac{N_{1j}\phi_j(\gamma)}{ N_{1j} \phi_j(\gamma) + N_{2j}} I_{21j}  \\
& \quad +  \frac{(n_{11j}+n_{21j})\phi_j(\gamma)}{ (N_{1j} \phi_j(\gamma) + N_{2j})^2 } (-N_{2j} I_{1j} + N_{1j} I_{2j}),
\end{align*}
which can be verified to be identical to $\hat h_j^{\mbox{\tiny LW}}(Y,\delta,Z; \gamma)$.
{\hfill $\Box$ \vspace{.1in}}

{\noindent \textbf{Proof of Proposition~\ref{pro:oddr-asym}.}}

The proof is similar to that of Proposition~\ref{pro:oddr-asym}.
First, it can be shown that $\hat\beta^{(w)}$ converges in probability to $\bar \beta^{(w2)}$ solving (\ref{eq:oddr-target}),
and $N^{1/2} ( \hat\beta^{(w)} - \bar \beta^{(w2)})$ converges in distribution to $\N(0, \Sigma^{(w2)})$, with
$ \Sigma^{(w2)}$ in the stated sandwich form, where $H^{(w2)}(\beta)$ is defined in Proposition~\ref{pro:oddr-asym}, but $G^{(w2)}(\beta)$ is defined such that
\begin{align*}
\frac{1}{N^{1/2}} \sum_{j=1}^J \frac{n_{11j} n_{22j} - \psi_j n_{12j} n_{21j}} { N_{1j} \psi_j + N_{2j}} x_j \to \N(0, G^{(w2)}(\beta) ).
\end{align*}
Here and subsequently, $\psi_j=\psi_j(\beta)$ and $\beta$ is evaluated at $\bar\beta^{(w2)}$. To complete the proof, it suffices to show that
\begin{align*}
& \frac{1}{N} \sum_{j=1}^J \frac{n_{11j} n_{22j} - \psi_j n_{12j} n_{21j}} { N_{1j} \psi_j + N_{2j}} x_j = \frac{1}{N} \sum_{i=1}^N  \sum_{j=1}^J  g_j(Y_i,\delta_i,Z_i;  \beta ) x_j + o_p(N^{-1/2}).
\end{align*}
Consider the decomposition using (\ref{eq:oddr-target}) and the previous decomposition (\ref{eq:decomp-MH}),
\begin{align*}
&\frac{1}{N} \sum_{j=1}^J \frac{n_{11j} n_{22j} - \psi_j n_{12j} n_{21j}} { N_{1j} \psi_j + N_{2j}} x_j \\
&= \frac{1}{N} \sum_{j=1}^J \frac{N_{2j} (p_{22j} +  \psi_j p_{21j}) (n_{11j} - p_{11j} N_{1j} )  - \psi_j N_{1j} (p_{11j} +  \psi_j p_{12j})(n_{21j} - p_{21j} N_{2j} )} { N_{1j} \psi_j + N_{2j}} x_j \\
& \quad + \frac{1}{N} \sum_{j=1}^J \frac{ (\psi_j-1) (n_{11j}- N_{1j}p_{11}) (n_{21j}-N_{2j} p_{21j}) } { N_{1j} \psi_j + N_{2j}} x_j \\
& \quad + \sum_{j=1}^J (p_{11j}p_{22j} - \psi_j p_{12j} p_{21j}) \left( \frac{N_{1j}N_{2j} /N}{N_{1j} \psi_j + N_{2j}} - \frac{P_{1j} P_{2j}}{ P_{1j} \psi_j + P_{2j}} \right) x_j.
\end{align*}
The first term can be shown to be
\begin{align*}
& \frac{1}{N} \sum_{j=1}^J \frac{N_{2j}  (p_{22j} +  \psi_j p_{21j})(n_{11j} - p_{11j} N_{1j} )  - \psi_j N_{1j}  (p_{11j} +  \psi_j p_{12j}) (n_{21j} - p_{21j} N_{2j} )} { N_{1j} \psi_j + N_{2j}} x_j \\
& = \frac{1}{N} \sum_{j=1}^J \frac{P_{2j} (p_{22j} +  \psi_j p_{21j}) (n_{11j} - p_{11j} N_{1j} )  - \psi_j P_{1j}  (p_{11j} +  \psi_j p_{12j}) (n_{21j} - p_{21j} N_{2j} )} { P_{1j} \psi_j + P_{2j}} x_j + o_p(N^{-1/2})  .
\end{align*}
The second term can be shown to be $o_p(N^{-1/2})$.
The third term can be shown to be
\begin{align*}
& \sum_{j=1}^J (p_{11j}p_{22j} - \psi_j p_{12j} p_{21j}) \left( \frac{N_{1j}N_{2j} /N}{N_{1j} \psi_j + N_{2j}} - \frac{P_{1j} P_{2j}}{ P_{1j} \psi_j + P_{2j}} \right) x_j \\
& = \sum_{j=1}^J (p_{11j}p_{22j} - \psi_j p_{12j} p_{21j}) \frac{(\frac{N_{1j}}{N}-P_{1j}) P_{2j}^2 + ( \frac{N_{2j}}{N} - P_{2j})\psi_j P_{1j}^2 }{ (P_{1j} \psi_j + P_{2j})^2} x_j + o_p(N^{-1/2})\\
& = \sum_{j=1}^J (p_{11j}p_{22j} - \psi_j p_{12j} p_{21j}) \frac{ \frac{N_{1j}}{N} P_{2j}^2 +  \frac{N_{2j}}{N}  \psi_j P_{1j}^2 }{ (P_{1j} \psi_j + P_{2j})^2} x_j + o_p(N^{-1/2}),
\end{align*}
where the last step follows because by (\ref{eq:oddr-target}),
\begin{align*}
& \sum_{j=1}^J (p_{11j}p_{22j} - \psi_j p_{12j} p_{21j}) \frac{ P_{1j}  P_{2j}^2 +  P_{2j} \psi_j P_{1j}^2 }{ (P_{1j} \psi_j + P_{2j})^2} x_j
 = \sum_{j=1}^J   (p_{11j}p_{22j} - \psi_j p_{12j} p_{21j}) \frac{P_{1j} P_{2j}}{ P_{1j} \psi_j + P_{2j}} x_j =0 .
\end{align*}
The desired expansion follows by collecting the preceding results.
{\hfill $\Box$ \vspace{.1in}}

\section{Additional numerical results}
\vspace{-.1in}

\begin{table}
\caption{Comparison of estimates from additional simulations of $2 \times 2$ tables} \label{tab:2by2-sim2}
\small
\begin{center}
\begin{tabular*}{.9\textwidth}{ll rrrr ll rrrr} \hline
     && Point   & SD       & bSE      & rSE     &&        & Point  &  SD  &  bSE      & rSE \\ \hline
     && \multicolumn{10}{c}{$(N_{1j},N_{2j}) = (30,20)$ and $J=4$} \\
MH   && $.7045$ & $.3381$  & $.3356$  & $.3370$ &&        &  \\
wMH  && $.7045$ & $.3381$  & $.3356$  & $.3370$ &&  BP    & $.2893$  & $.1410$ & $.1433$ & $.1422$ \\
CML  && $.7002$ & $.3357$  & $.3338$  & NA      && oldBP  & $.2893$  & $.1410$ & $.2145$ & NA \\   \cline{2-12}
     && \multicolumn{10}{c}{$(N_{1j},N_{2j}) = (3,2)$ and $J=40$} \\
MH   && $.7109$ & $.3466$  & $.3492$  & $.3525$ &&        &  \\
wMH  && $.7109$ & $.3466$  & $.3492$  & $.3525$ &&  BP    & $.3419$  & $.1684$ & $.1725$ & $.1720$ \\
CML  && $.7061$ & $.3426$  & $.3455$  & NA      && oldBP  & $.3419$  & $.1684$ & $.2391$ & NA \\  \hline
\end{tabular*} \\[1ex]
\parbox{.9\textwidth}{\scriptsize Note: Log odds ratio $=\log(2) = .6931$.}
\end{center} \vspace{-.1in}
\end{table}

Table~\ref{tab:2by2-sim2} presents simulation results from two settings where the probabilities $(p_{11j},p_{21j})$ are spread out in $(0,1)$.
For the first setting, $J=4$ tables are simulated with log odds ratio $\psi_j = \log(2)$,
probabilities $p_{11j} = .05 + .2j$ between $.25$ and $.85$ for $j=1,\ldots,4$, and binomial sizes $(N_{1j},N_{2j})= (30, 20)$.
This corresponds to asymptotic Setting I (large tables).
For the second setting, $J=40$ tables are simulated with log odds ratio $\psi_j = \log(2)$,
probabilities $p_{11j} = .05 + .02j$ between $.07$ and $.85$ for $j=1,\ldots,40$, and binomial sizes $(N_{1j},N_{2j})= (3, 2)$.
This corresponds to asymptotic Setting II (many sparse tables).
In these two settings with common odds ratios and constant $(N_{1j},N_{2j})$ in $j$, the weighted Mantel--Haenszel estimator reduces to the unweighted Mantel--Haenszel estimator.
The model-based and model-robust variance estimators for wMH and CML reasonably match the Monte Carlo standard deviations.
The BP point estimator, being centered around a target probability ratio, is smaller than the true log odds ratio $\log(2)$.
The model-based variance estimator on the row BP agrees with the Monte Carlo standard deviation, but
the commonly reported variance estimator on the row oldBP is biased upward.

\begin{figure}
\begin{tabular}{c}
\includegraphics[width=6.2in, height=2.5in]{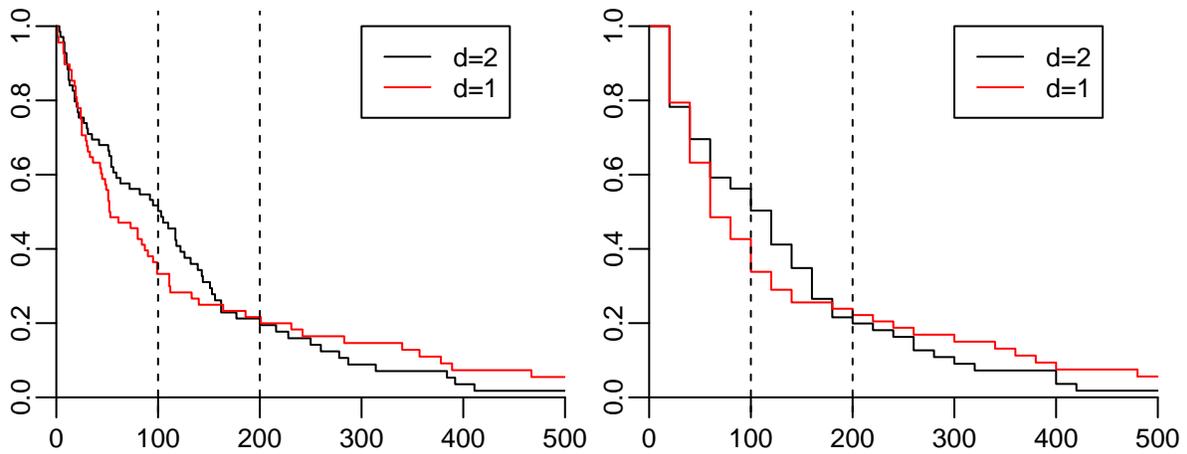}
\end{tabular}
\caption{Kaplan--Meier survival curves for the original (left) and discretized data (right) from Veteran's lung cancer trial.}
\label{fig:VA} \vspace{.1in}
\end{figure}

Figure~\ref{fig:VA} shows the Kaplan--Meier survival curves in the two groups for the Veteran's lung cancer data and discretized data.
The two survival curves cross each other, indicating
non-proportional hazards over time between the two groups.

\begin{figure}
\begin{tabular}{c}
\includegraphics[width=6.2in, height=2.5in]{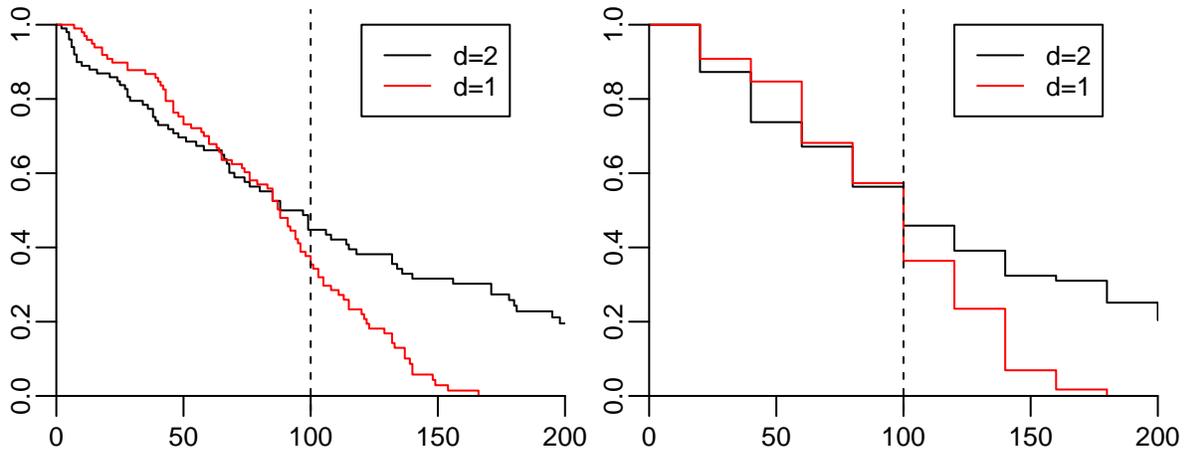}
\end{tabular}
\caption{Kaplan--Meier survival curves for finely (left) and coarsely (right) discretized data from Weibull distributions. The time axis is rescaled by 100.}
\label{fig:weibull} \vspace{.1in}
\end{figure}

\begin{figure}
\begin{tabular}{c}
\includegraphics[width=6.2in, height=2.5in]{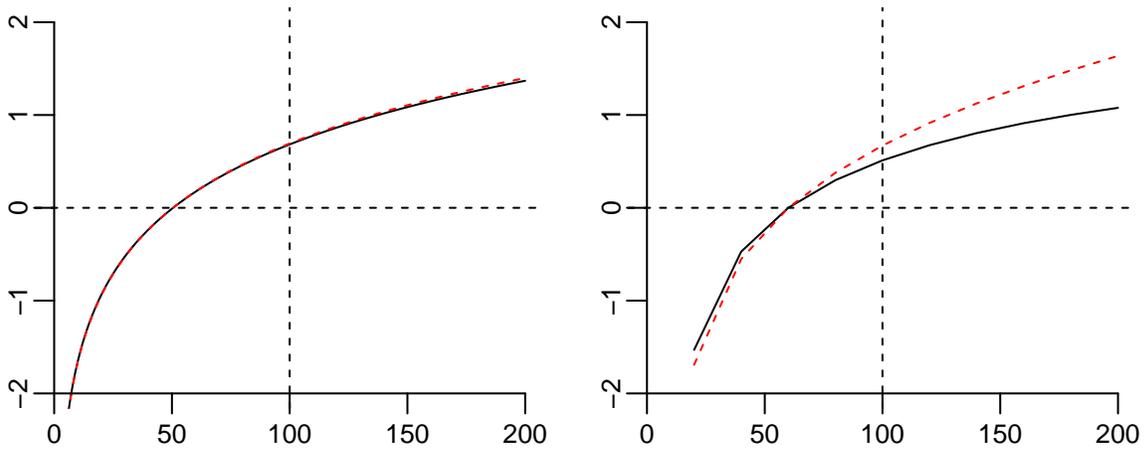}
\end{tabular}
\caption{Log probability ratios (solid) and log odds ratios (dashed) over time for the two Weibull distributions used in the simulations, finely (left) and coarsely (right) discretized. The time axis is rescaled by 100.}
\label{fig:weibull2} \vspace{.1in}
\end{figure}

Figure~\ref{fig:weibull} shows the Kaplan--Meier survival curves for finely and coarsely discretized data, simulated from Weibull distributions.
Figure~\ref{fig:weibull2} shows the true log probability and odds ratios over time from the two Weibull distributions.
The log probability and odds ratios are virtually the same for finely discretized data, but the log probability ratios are
closer to 0 than log odds ratios for coarsely discretized data.

\vspace{.4in}
\centerline{\bf\Large Supplement References}

\begin{description}
\item Flanders, W.D. (1985) A new variance estimator for the Mantel--Haenszel odds ratio, {\em Biometrics}, 41, 637--642.

\item Tan, Z. (2004) On a likelihood approach for Monte Carlo integration, {\em Journal of the American Statistical Association}, 99, 1027--1036.
\end{description}

\end{document}